\documentclass[aps,preprint]{revtex4}
\usepackage{xcolor}
\usepackage{amsfonts}
\usepackage{amssymb}
\usepackage{amsmath}
\usepackage[font={footnotesize,it}]{caption}
\usepackage{latexsym}
\usepackage{changes}
\usepackage{hyperref}
\usepackage[capitalise]{cleveref}

\usepackage[caption = false]{subfig}
\usepackage{graphicx}
\usepackage{float}
\usepackage[font=small,labelfont=bf,
   justification=justified,
   format=plain]{caption}
\hypersetup{
    colorlinks=true,
    linkcolor=blue,
    filecolor=magenta, citecolor=blue,    urlcolor=blue,
}
\begin{document}

\title{The Charged Zipoy - Voorhees Metric with Astrophysical Applications}
\author{Ozay Gurtug}
\email{ozaygurtug@maltepe.edu.tr}
\affiliation{T. C. Maltepe University, Faculty of Engineering and Natural Sciences,
34857, Istanbul -Turkey}
\author{Mustafa Halilsoy}
\email{mustafa.halilsoy@emu.edu.tr}
\affiliation{Department of Physics, Faculty of Arts and Sciences, Eastern Mediterranean
University, Famagusta, North Cyprus via Mersin 10, Turkey}
\author{Mert Mangut}
\email{mert.mangut@emu.edu.tr}
\affiliation{Department of Physics, Faculty of Arts and Sciences, Eastern Mediterranean
University, Famagusta, North Cyprus via Mersin 10, Turkey}

\begin{abstract}
Starting from an integral of the interaction region of colliding Einstein-Maxwell  waves and by applying a coordinate transformation, we obtain the charged version of the static Zipoy-Voorhees $(ZV)$  metric valid for all values of the distortion parameter $\gamma$. In Schwarzschild coordinates, we investigate the effect of the charge in the newly found spacetime, stress the analogy with Reissner - Nordstrom metric and discuss some of its features. It is shown that from the expression of Weyl curvature, directional  singularities become manifest. For astrophysical importance, we find lensing of null geodesics from the Gauss - Bonnet theorem in such non - spherically charged objects. To prepare the ground for our null, circular geodesics we consider the angular equation linearized about the symmetry plane $\theta=\pi/2$. This, in turn, suggests the distortion parameter (the $ZV$ parameter) must be in the interval $1/2<\gamma<1$. It is found that the lensing angle is highly dependent on the distortion parameter, and becomes decisive on the effect of the charge. For a class of charged compact stars, we plot the deflection angle versus the ratio of impact parameter to the radius of the star. Plots have revealed that for  perfectly spherical compact stars, it is hard to identify the effect of electric/magnetic charge, but for non-spherical compact stars the effect of electric charge becomes apparent. For comparison, the same lensing angle has also been found for the stationary $ZV$ metric in the equatorial plane. Our analysis indicates that depending on the value of $\gamma$, the stationary state induces counter effect on the bending angle and thus, when compared with the uncharged static $ZV$ case, the bending angle decreases. The influence of the parameter $\gamma$ on the gravitational redshift is also displayed.

\end{abstract}

\maketitle

\section{Introduction}
The existence of non-spherical stellar objects brings into focus the geometry of metrics with intricate topologies known as the $\gamma-$ metric or Zipoy-Voorhees $(ZV)$  class of metrics \cite{1,2,3}. The geometry of these non-spherical topologies naturally affects particles/satellites and light in a non-isotropic manner, depending on the magnitude of the $\gamma-$ parameter $(0<\gamma<\infty)$. The $\gamma=1$ corresponds to a spherical object which affects all particles  passing by the source equally. For $\gamma>1$ and  $\gamma<1$, however, the deflection of all geodesics particles shows great variations as depicted for the null geodesics. In the astrophysical application, we have shown explicitly the variation of light deflection with the changing $\gamma$. As a result, the role of the $\gamma-$ factor in the metric becomes rather important and from the deflection data of light, it is possible to classify the stellar objects accordingly. This constitutes the main motivation for the present paper. It is also known that a spinning source automatically diverts from spherical symmetry and the spinning shows itself in the surrounding spacetime as in the case of rotating black holes. In the present study, we shall concentrate mainly on non-spinning objects with restricted deformation parameter $\gamma$. Exception to this restriction will be the equatorial plane for a specific stationary $ZV$-spacetime. The reason for the limitation on the $\gamma$ parameter originates from the fact that when we come to the geodesic analysis the nonintegrability restricts the deformation parameter to the range $1/2<\gamma<1$. We have used the data of a class of compact stars to show explicitly that increasing  $\gamma$ causes an increase in the deflection of light passing by the compact star. Besides investigating the role of $\gamma$, we also take into account the role of charge on the source which further affects the amount of deflection. In particular, strong magnetic fields of any stellar object are known to make a big difference. Although there is a growing interest in such planetary objects when searched in the literature, for their charged version, one comes accross with only a few references \cite{4,5}. In their approach \cite{4}, test field solution in a background geometry is found by the Killing symmetry and from the energy-momentum of the resulting electromagnetic $(em)$ field, the authors proceed to establish the Einstein-Maxwell $(EM)$ spacetime. Expectedly, the final part is technically difficult and for that reason, we follow an alternative procedure in this article. For the same purpose, we approach the problem from the spacetime of colliding $(EM)$ waves formulated long ago \cite{6,7} and transform the metric into the static, non-spherical form of $ZV$. In certain sense, we apply the principle of holography to the physics of colliding plane waves in order to obtain a physical problem of $3-$ dimensions. More clearly, the field equations are solved in prolate type coordinates as a mathematical tool which are found to be very useful in the description of colliding gravitational waves. Furthermore, the resulting solution is interpreted in some other coordinates. For this purpose, we use the line element of Chandrasekhar and Xanthopoulos $(CX)$ \cite{8,9} as the starting point and employ the method of \cite{10}. The spacetime of colliding $EM$ waves is described by the line element

\begin{equation}
ds^{2}=\sqrt{\Delta}e^{N} \left( \frac{d\eta^{2}}{\Delta}-\frac{d\mu^{2}}{\delta}  \right)-\sqrt{\Delta \delta}\left( \chi dx^{2} + \chi^{-1} dy^{2}  \right)
\end{equation}
in which the metric functions $(N,\chi)$ depend on $(\eta,\mu)$ alone, $\Delta=1-\eta^{2}$ and $\delta=1-\mu^{2}$. The coupled Ernst equation \cite{11,12} for $EM$ fields are given by the pair

\begin{equation}
\left(\xi\bar{\xi}+\eta\bar{\eta}-1 \right)\nabla^{2}\xi=2\nabla\xi \left( \bar{\xi}\nabla\xi+\bar{\eta}\nabla\eta \right)
\end{equation}

\begin{equation}
\left(\xi\bar{\xi}+\eta\bar{\eta}-1 \right)\nabla^{2}\eta=2\nabla\eta \left( \bar{\xi}\nabla\xi+\bar{\eta}\nabla\eta \right)
\end{equation}
where $\xi$ and $\eta$ represent the gravitational and $em$ complex potentials, respectively. Note that the bar denotes complex conjugation and the operators $\nabla$ and $\nabla^{2}$ are defined on an appropriate base manifold. We also record  that an alternative parametrization for the Ernst potentials are

\begin{equation}
Z=\frac{1+\xi}{1-\xi}
\end{equation}

\begin{equation}
H=\frac{\eta}{1-\xi}
\end{equation}
where $Z$ and $H$ become the gravitational and $em$ potentials, respectively. Further parametric transformations on $(Z,H)$ functions are known to generate new solutions, especially metrics with cross terms \cite{19}. Our  interest in this  work will be confined only to the diagonal metrics with elaboration on the role of the distortion parameter and electric/magnetic charge in astrophysical objects. The emergence of angular dependent singularities is shown explicitly with the addition of charge and distortion parameters. The rest of the paper is organized as follows. In   section \ref{sect2},  we  shall solve the Ernst equation, transform the metric into Schwarzschild coordinates and in section \ref{sect3}, we shall discuss its main properties. A restricted geodesic analysis to justify  our choice of $\theta=\pi/2$ appears in section \ref{sect4}. In section \ref{sect5}, we discuss lensing in the charged $ZV$ and uncharged stationary ZV metrics and gravitational redshift effect for charged $ZV$ with astrophysical applications in section \ref{sect6}.  Our conclusion will be presented in section \ref{sect7}.

\section{Solution for the Ernst system}\label{sect2}

A particular, real solution to the foregoing set of equations $(2,3)$, is given by

\begin{equation}
\xi=p\xi_{0},
\end{equation}
and

\begin{equation}
\eta=q\xi_{0},
\end{equation}
where $\xi_{0}=\bar{\xi_{0}}$ and the real constants $(p,q)$ satisfy

\begin{equation}
p^{2}+q^{2}=1.
\end{equation}

Clearly, the parameter $q$ is a measure of charge. The choice $\xi_{0}=tanhX$, where $X(\eta,\mu)$ satisfies the Euler-Darboux equation

\begin{equation}
\left( \Delta X_{,\eta} \right)_{,\eta}-\left( \delta X_{,\mu} \right)_{,\mu}=0,
\end{equation}
in which a comma denotes partial derivative, solves the Ernst system in the real domain. We now make the choice

\begin{equation}
e^{2X}=\left(\frac{1-\eta}{1+\eta} \right)^{\gamma},
\end{equation}
where $\gamma$ is a constant to be identified in the sequel as the $ZV-$ parameter. From the references \cite{8,9,10}, it is known that the metric function $\chi$ is given by

\begin{equation}
\chi=\frac{\sqrt{\Delta\delta}}{\Psi},
\end{equation}
where the function $\Psi$ is determined from

\begin{equation}
Z=\Psi + H^{2}.
\end{equation}

The remaining metric function $N$ is obtained from the integrability equations \cite{10}

\begin{equation}
\left(N+ln\Psi \right)_{,\eta}=\frac{2\eta}{\delta-\Delta}+\frac{\eta}{\Delta}+\frac{2\delta}{\delta-\Delta}\left[2\mu\Delta X_{,\eta}X_{,\mu}-\eta \left( \Delta X^{2}_{,\eta}+ \delta X^{2}_{,\mu}  \right)    \right].
\end{equation}

\begin{equation}
\left(N+ln\Psi \right)_{,\mu}=\frac{2\mu}{\Delta-\delta}+\frac{2\Delta}{\Delta-\delta}\left[2\eta\Delta X_{,\eta}X_{,\mu}-\mu \left( \Delta X^{2}_{,\eta}+ \delta X^{2}_{,\mu}  \right)    \right].
\end{equation}

Upon substitution for $X$ from $(10)$ and integrating for the metric function  $N$, we obtain the final line element

 \begin{equation}
ds^{2}=M^{2}(\eta) \left[ \Delta^{\gamma^{2}}\left( \delta-\Delta \right)^{1-\gamma^{2}}\left( \frac{d\eta^{2}}{\Delta}-\frac{d\mu^{2}}{\delta}  \right)-\Delta\delta dx^{2} \right]-\frac{dy^{2}}{M^{2}(\eta)},
\end{equation}
in which

 \begin{equation}
M(\eta)=coshX-psinhX.
\end{equation}

At this point, it is worthwile to add that this line element  must satisfy the appropriate boundary conditions of the incoming and interaction regions in order to be considered as a solution to the problem of colliding $EM$ waves. In this regard, $(15)$ is not a promising candidate as the factor $(\delta-\Delta)$ in the metric fails to satisfy the boundary conditions. As a solution, however, in the interaction region alone it is useful, as we shall show below.

Having applied the coordinate transformation

\begin{eqnarray}
p\eta+1=\frac{r}{m} \;\;\; ,  \text{ \
\ \ }x=\varphi \notag, \\
\mu=cos\theta   \;\;\;,  \text{ \
\ \ }y=\tau,
\end{eqnarray}
to the line element $(15)$, supplemented with the condition $(-1)^{\gamma}=-1$, and appropriate rescaling of time $(\tau \rightarrow t)$, we cast the metric into the form

 \begin{equation}
ds^{2}=\frac{\Delta^{\gamma}}{K^{2}}dt^{2}-\frac{K^{2}}{\Delta^{\gamma}}\left[ \Delta^{\gamma^{2}}\Sigma^{1-\gamma^{2}}\left( \frac{dr^{2}}{\Delta}+r^{2}d\theta^{2}  \right)+r^{2}\Delta sin^{2}\theta d\varphi^{2}  \right].
\end{equation}

Our notation here is as follows

 \begin{equation}
\Delta(r)=1-\frac{2m}{r}+\frac{m^{2}q^{2}}{r^{2}},
\end{equation}

\begin{equation}
\Sigma(r,\theta)=1-\frac{2m}{r}+\frac{m^{2}}{r^{2}}\left( q^{2}+p^{2}sin^{2}\theta \right),
\end{equation}
and

\begin{equation}
K(r)=(1+p)\left(1-\frac{m(1-p)}{r}   \right)^{\gamma}-(1-p)\left(1-\frac{m(1+p)}{r}   \right)^{\gamma}.
\end{equation}

It is checked that  line element $(18)$ solves the $EM$ equations and for $q=0$ $(p=1)$, it reduces to the $ZV-$ metric. It is seen also that the case $p=0$ $(q=1)$ must be excluded since the metric function $K=0$ in such a limit. Furthermore, although we assumed the condition $(-1)^{\gamma}=-1$ for the parameter $\gamma$ in the begining, we noticed that this condition can be released and the metric becomes valid for all $\gamma $'s. The reason is that the transformation for $\gamma=even$ also can be incorporated through an analytic continuation $y\rightarrow i\tau$ and $x \rightarrow i \varphi$, letting the other coordinates as in the previous  transformation. We remind that charged version of similar metrics were given before \cite{20}, which were restricted only to the integer parameters. The new solution $(18)$ represents charged, deformed $ZV-$ objects that is valid for all $\gamma$'s. Since planetary objects are mostly charged, especially magnetic, this metric will have astrophysical applications.  In the next section, we investigate some of the features of our new metric $(18)$.

\section{Properties of the solution}\label{sect3}

In this section, we shall investigate some of the properties of our metric $(18)$. Firstly, we shall expand the time component of the metric to see the asymptotic behaviour for large $r$ values. This will show the Newtonian potential in the presence of both mass and charge. Secondly, we show that static magnetic and electric fields can separately  be considered as the source to our spacetime. And thirdly, we shall consider the singularity distribution in our metric.

\subsection{ Newtonian Limit}

As usual, in order to see the Newtonian limit, we take the $g_{tt}$ component of the metric and express it in the form

\begin{equation}
g_{tt}=\frac{\Delta^{\gamma}}{K^{2}} \approx 1+2\phi,
\end{equation}
where $\phi=\phi(r)$ is the asymptotic expression of the Newtonian potential. Upon expansion to the order $r^{-3}$, we obtain

\begin{equation}
\phi(r) \approx -\frac{m\gamma}{r}+\frac{\gamma m^{2}}{2r^{2}}\left[ \gamma q^{2}+2(\gamma-1) \right]-\frac{\gamma(\gamma-1)}{3}\left(\frac{m}{r}\right)^{3}\left[ (1+4\gamma)q^{2}+2(\gamma-2)  \right]+\mathcal{O}\left( \frac{1}{r^{4}}\right).
\end{equation}

It is observed that the physical mass, $M=m\gamma$ and  charge $Q=mq$ play roles at the monopole, dipole and the quadrupole orders. It will be in order at this point to comment that without charge $(q=0)$, the dipole term $\sim 1/r^{2}$  can be removed by a transformation on the radial coordinate \cite{18}. We shall ignore such a procedure since it will be a repetition. Since we have charge in the present case, the $1/r^{2}$ term can be attributed to the charge. For the simplest case of spherical symmetry $(\gamma=1)$, we recover the asymptotic form of the Reissner-Nordrstrom $(RN)$ potential. It is observed also that the case of charge without mass is not available. However, for the pure electromagnetic limit in the case of $\gamma=1$, which yields the Bertotti-Robinson $(BR)$ metric \cite{13,14}, one must consult \cite{10}.

\subsection{Electromagnetic Sources for the Metric}

\subsubsection{Pure magnetic case}

The vector potential is chosen as

\begin{equation}
A_{\mu}=(0,0,0,C_{0}cos\theta),
\end{equation}
where $C_{0}$  is the magnetic charge proportional to $q$. The magnetic field becomes $F_{\theta\varphi}=-C_{0}sin\theta$ from $F_{\mu\nu}=\partial_{\mu}A_{\nu}-\partial_{\nu}A_{\mu}$, which solves the only relevant source-free Maxwell equation

\begin{equation}
\nabla_{\mu}F^{\mu\nu}=0.
\end{equation}

The invariant of the magnetic field is given by

\begin{equation}
I=F_{\mu\nu}F^{\mu\nu}=\frac{2C^{2}_{0}}{r^{4}K^{4}}\Delta^{-(\gamma-1)^{2}}\Sigma^{\gamma^{2}-1},
\end{equation}

which trivially reduces to the case of $RN$ for $\gamma=1$. Any divergence in the electromagnetic field can easily be identified from this invariant, which is highly dependent on $\gamma$. Directional singularities can also be identified for the $em$ field from the zeros of $\Sigma$ for certain $\gamma$ values $(\gamma^{2}<1)$.

It is interesting to see that although the $em$ invariant diverges at $r=0$; when $\gamma=1$,  it becomes regular for $\gamma \neq 1$. To see this, we express $K$ in the form $K=r^{-\gamma}K_{0}$ and upon substituting $r=0$, the invariant becomes $I=\frac{2C^{2}_{0}}{K_{0}^{4}}(m^{2}q^{2})^{-(\gamma-1)^{2}}(m^{2}q^{2}+m^{2}p^{2}sin^{2}\theta)^{\gamma^{2}-1}<\infty$ since $K_{0}\neq 0$. We also add  that from the Einstein equations $R_{\mu\nu}=-T_{\mu\nu}=\frac{1}{4}g_{\mu\nu}F_{\alpha\beta}F^{\alpha\beta}-F_{\mu\alpha}F_{\nu}^{\alpha}$, the constant $C_{0}$ can be fixed as $C_{0}^{2}=8m^{2}\gamma^{2}p^{2}q^{2}$ so that $T_{\mu}^{\nu}=\frac{I}{4}diag(-1,-1,1,1)$. Although this solution does not correspond to a magnetic dipole, we recall that at large distances the magnetic charge $q$, from the expansion $(23)$ multiplied by the Legendre polynomial $P_{1}=cos\theta$ plays the role of a dipole.

\subsubsection{Pure electric case}

In this case, we choose the vector potential as

\begin{equation}
A_{\mu}=(f(r),0,0,0),
\end{equation}
where $f(r)$ is a function of $r$ which is to be determined from the satisfaction of the Maxwell equation $\nabla_{r}F^{rt}=0$. Up to a trivial additive constant, we obtain

\begin{equation}
f(r)=C_{1} \int^{r} \frac{\Delta^{\gamma-1}dr}{r^{2}K^{2}},
\end{equation}
where $C_{1}$  is  an  integration constant propartional to $q$. It is seen that finding the exact form of $f(r)$ depends entirely on the $\gamma-$ term and this expression is given explicitly in the sequel. By choosing the spherical symmetric case $\gamma=1$, one easily observes that  the potential of the pure electric $RN$ solution is recovered.

The invariant of the field is computed in this case as

\begin{equation}
I=F_{\mu\nu}F^{\mu\nu}=-\frac{2C^{2}_{1}}{r^{4}K^{4}}\Delta^{-(\gamma-1)^{2}}\Sigma^{\gamma^{2}-1},
\end{equation}
which is in conform ( up to an expected sign change) with the magnetic invariant. Just as the magnetic case, the constant $C_{1}$ is  found as $C_{1}^{2}=8m^{2}\gamma^{2}p^{2}q^{2}$ and the energy-momentum tensor is $T_{\mu}^{\nu}=\frac{I}{4}diag(1,1,-1,-1).$ If we put the constant $C_{1}$ into integral (28) and integrate, the electrical potential can be written as

\begin{equation}
A_{0}(r)=\frac{\sqrt{2}q\left(1-\frac{m(1+p)}{r} \right)^{\gamma}}{(1+p)K(r)}.
\end{equation}

The variation of electric potential for specific charge related parameter $q$ against $r$ is plotted for different values of the deformation parameter $\gamma$. From the figures, we observed that the effect of the parameter $\gamma$ becomes weaker towards the asymptotic infinity. However, the effect of $\gamma$ becomes more apparent in the near regions.

\begin{figure}[H]
\centering
  \begin{tabular}{@{}cccc@{}}
    \includegraphics{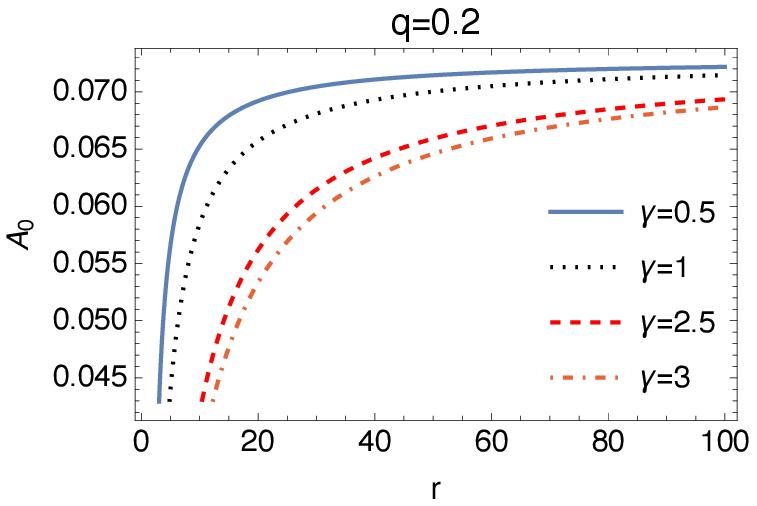} &
    \includegraphics{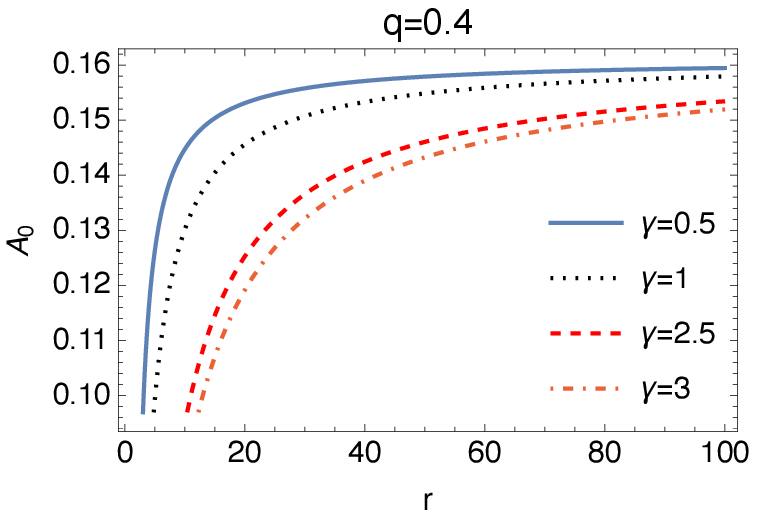} &
                                                         \\
    \includegraphics{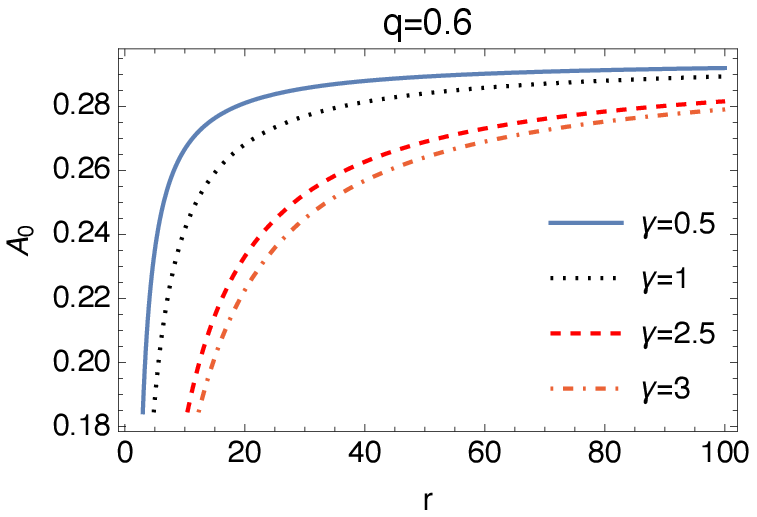} &
    \includegraphics{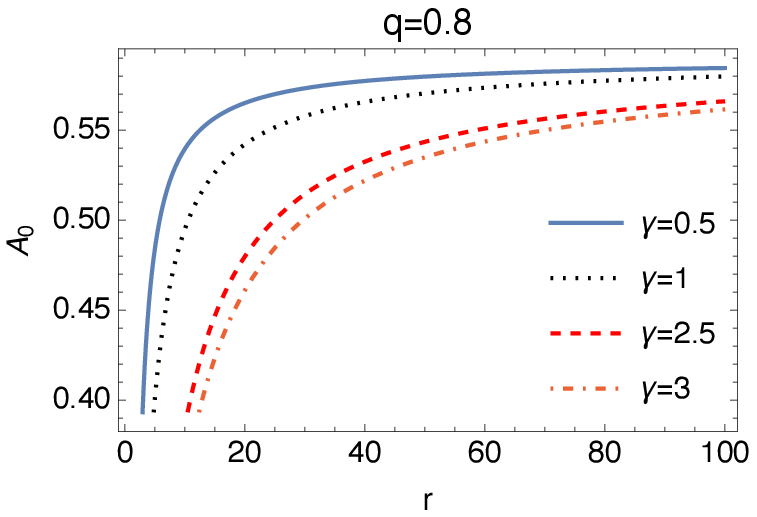} &
                                                            \\
    \end{tabular}
  \caption{Radial dependence of the electric potential in the charged ZV metrics. Plots are generated for $m=1$ and different values of $\gamma$ and charge related parameter $q$.}
\end{figure}

\subsection{ Singularity Detection}

Our metric contains singularities. In order to see this we must compute the Kretchmann scalar $R_{\mu\nu\alpha\beta}R^{\mu\nu\alpha\beta}$, however, due to the technical difficulties we follow an alternative way to investigate the Newman-Penrose $(NP)$ component \cite{15}  $\Psi_{2}$ of null-tetrad formalism. In general, the singularities of $\Psi_{2}$ are indicative to the singularities of the overall spacetime. It will be seen that even this will not be an easy task at all. For this purpose, we make the choice of null-tetrad basis $1-$ forms

\begin{equation}
\begin{aligned}
\sqrt{2}l&=A(r)dt-C(r)sin\theta d\varphi=\sqrt{2}l_{\mu}dx^{\mu}, \\
\sqrt{2}n&=A(r)dt+C(r)sin\theta d\varphi=\sqrt{2}n_{\mu}dx^{\mu}, \\
\sqrt{2}m&=B(r,\theta)\left(\frac{dr}{\sqrt{\Delta}}+ird\theta \right)=\sqrt{2}m_{\mu}dx^{\mu}, \\
\sqrt{2}\bar{m}&=B(r,\theta)\left(\frac{dr}{\sqrt{\Delta}}-ird\theta \right) =\sqrt{2}\bar{m}_{\mu}dx^{\mu},
\end{aligned}
\end{equation}
where $A(r)$, $B(r,\theta)$  and  $C(r)$  are abbreviated as

\begin{equation}
\begin{aligned}
A(r)&=\Delta^{\gamma/2}K^{-1},\\
C(r)&=Kr\Delta^{\frac{1-\gamma}{2}}, \\
B(r,\theta)&=K\Delta^{\frac{\gamma^{2}-\gamma}{2}} \Sigma^{\frac{1-\gamma^{2}}{2}}.
\end{aligned}
\end{equation}

In the null-tetrad $(l_{\mu},n_{\mu},m_{\mu},\bar{m}_{\mu})$ of $NP$ we find the following expression for $\Psi_{2}$;

\begin{equation}
\begin{aligned}
\Psi_{2}(r,\theta)=&\frac{\Delta^{1+\gamma-\gamma^{2}}}{4K^{2}}\Sigma^{\gamma^{2}-1}\left\{\frac{\Delta^{\prime}}{2r\Delta}- 3\left(\frac{K^{\prime}}{K}\right)^{2}+\frac{1}{4}\gamma(\gamma-1)(\gamma-\gamma^{2}-2)\left(\frac{\Delta^{\prime}}{\Delta}\right)^{2}  \right.\\
&\left.+\frac{1}{4}(1-\gamma^{2})(\gamma^{2}-2)\left(\frac{\Sigma^{\prime}}{\Sigma}\right)^{2}+\gamma(2-\gamma)\frac{K^{\prime}}{K}\frac{\Delta^{\prime}}{\Delta}+(\gamma^{2}-1)\frac{K^{\prime}}{K}\frac{\Sigma^{\prime}}{\Sigma}
\right.\\
&\left. +\frac{1}{4}(1-\gamma^{2})(1+2\gamma-2\gamma^{2})\frac{\Sigma^{\prime}}{\Sigma}\frac{\Delta^{\prime}}{\Delta}-\frac{2}{r}\frac{K^{\prime}}{K}+\frac{\gamma^{2}}{2r}\frac{\Delta^{\prime}}{\Delta}+\frac{K^{\prime\prime}}{K} +\frac{1}{2}\gamma(\gamma-1)\frac{\Delta^{\prime\prime}}{\Delta}
\right.\\
&\left.+\frac{1}{2}(1-\gamma^{2}) \frac{\Sigma^{\prime\prime}}{\Sigma}+\frac{(1-\gamma^{2})p^{2}m^{2}}{r^{4}\Sigma^{2}}\left( cos2\theta-\frac{p^{2}m^{2}}{\Delta r^{2}}sin^{2}\theta  \right) \right\},
\end{aligned}
\end{equation}
in which  prime denotes derivative with respect to $r$ of the functions given in $(18)$.

We observed that further substitution of the derivatives will add little other than the already wild expression for the Weyl curvature. Even the study of this expression for $\Psi_{2}(r,\theta)$ reveals the occurrence of singularities at $r=0$, and the roots of $\Delta(r)=0$ and $\Sigma(r,\theta)=0$. It is interesting to observe that the outermost singularity is due to the root of $\Delta(r)=0$, which gives

\begin{equation}
r_{\Delta}=m(1+p).
\end{equation}

The larger root of $\Sigma(r,\theta)=0$, gives

\begin{equation}
r_{\Sigma}=m(1+pcos\theta),
\end{equation}
which lies inside of $r_{\Delta}$. Thus, any probe to the singularities of the metric must encounter first with $r_{\Delta}=m(1+p)$. Another point of interest is to comment that for $r>0$ and $0<p<1$, we have $K(r)\neq 0$, so that no additional singularity arises due to the presence of $K(r)$ in the expression for curvature. We remark that the function $K(r)$ is the metric function that arises when Maxwell field is added to the gravitational $ZV-$ metric. $K(r)$ becomes a constant when the electromagnetic source vanishes. A final observation from $\Psi_{2}(r,\theta)$ about singularities is that directional singularities occur for $\gamma^{2}<3$, since beyond this interval the power of $\Sigma(r,\theta)$ becomes positive to avoid  any divergence.

\section{Restricted Geodesic Analysis for the $ZV$ Spacetime}\label{sect4}

We note that a general geodesic analysis reqires a separate study in its own right which is beyond our scope. Our intension is to show that we can consider null-circular geodesics - at least - for the particular case of $\theta=\pi/2$. With this restriction we shall lose the generality of the $\gamma-$ parameter, which will be confined upon the discussion in this section to $1/2<\gamma<1.$ Firstly, we shall show the contribution of charge to the time-like circular geodesics. Next, we shall discuss the circular null-geodesic for $\dot{r}=0$, $\dot{\theta}\neq 0$ and thirdly the circular null-geodesics with $\dot{r}=0$, $\dot{\theta}\neq 0$ in linear approximation.

\subsection{Charge Effect on the Time-Like Geodesics with $\dot{r}=\dot{\theta}=0$}

The spacetime line element reduces under these restrictions to

 \begin{equation}
ds^{2}=A(r)dt^{2}-r^{2}C_{0}(r)sin^{2}\theta d\varphi^{2},
\end{equation}
where $A(r)$ and $C_{0}(r)$ can be identified from the general from of the metric (18). A reduced Lagrangian to describe the system is

 \begin{equation}
\mathcal{L}=\frac{1}{2}A\dot{t}^{2}-\frac{1}{2}r^{2}C_{0}(r)sin^{2}\theta\dot{\varphi}^{2},
\end{equation}
in which a dot represents derivative with respect to an affine parameter. For the time-like circular geodesics we have the angular velocity squared

 \begin{equation}
\omega^{2}=\frac{\dot{\varphi}^{2}}{\dot{t}^{2}}=\frac{A^{\prime}}{r^{2}C_{0}^{\prime}sin^{2}\theta},
\end{equation}
where a prime means derivative with respect to $r$. Asymptotic expansion of each term, with the choice $K \approx1$, upon scaling ( and detailed expansions that we present in the next section ) results in

 \begin{equation}
\omega^{2} \approx \frac{m \gamma}{r^{3}sin^{2}\theta} \left( 1+\frac{3m}{r}(1-\gamma)+\frac{m\gamma}{r}q^{2} +...\right).
\end{equation}

We recall that for $\gamma=1$ and $\theta=\pi/2$ we recover that Kepler's orbital law. For $q=0$ this result agrees with the result of chargeless, circular geodesics \cite{23}. It is seen that for $\gamma=1$, i.e. the Reissner-Nordstrom case we have the charge contribution at the order of $ 1/r^{4}$. We wish to draw attention that this result is valid also for $\theta=\pi/2.$

\subsection{The Circular Null-Geodesics}

In this particular case we take the reduced line element (36) to vanish which implies that the square of the angular frequency  is

 \begin{equation}
\omega^{2}=\frac{\dot{\varphi}^{2}}{\dot{t}^{2}}=\frac{A}{r^{2}C_{0}sin^{2}\theta}.
\end{equation}

Upon substitution from the metric (18) we obtain

 \begin{equation}
\omega^{2}=\frac{1}{r^{2}K^{4}sin^{2}\theta}\left(1-\frac{2m}{r}+\frac{m^{2}q^{2}}{r^{2}} \right)^{2\gamma-1}
\end{equation}
or equivalently

 \begin{equation}
\omega^{2}=\frac{(r-r_{1})^{2\gamma-1}}{r^{4\gamma}K^{4}sin^{2}\theta}\left(r-r_{\Delta} \right)^{2\gamma-1}
\end{equation}
where $r_{1}=m(1-p)$ and $r_{\Delta}=m(1+p)$, from (34). It is seen that for $p=1$, $q=0$, the results reduce to that of chargeless $ZV$. This suggests that in order to define $\omega^{2}$ as meaningful we must have $\gamma>1/2$. For $\gamma<1/2$, $\omega^{2}$ diverges at the outer root $r=r_{\Delta}$ and we must avoid this. We remark that this conclusion is valid for $\theta\neq\pi/2$ as well as for $\theta=\pi/2$.

\subsection{The Linearized Circular Geodesics with $\dot{r}=0$, in the vicinity of $\theta=\pi/2$}

The geodesic Langrangian can be chosen now in the form

\begin{equation}
\mathcal{L}=\frac{1}{2}A\dot{t}^{2}-\frac{1}{2}r^{2}B(r,\theta)\dot{\theta}^{2}-\frac{1}{2}r^{2}sin^{2}\theta C_{0}(r)\dot{\varphi}^{2}.
\end{equation}

The Euler-Lagrange equations give

 \begin{equation}
\dot{t}=\frac{E}{A},
\end{equation}

 \begin{equation}
\dot{\varphi}=\frac{l}{C_{0}(r)r^{2}sin^{2}\theta},
\end{equation}
where $E$ and $l$ are integration constants. Once we impose the null condition $ds^{2}=0$, we obtain the constraint

 \begin{equation}
r^{2}B(r,\theta)\dot{\theta}^{2}=\frac{E^{2}}{A}-\frac{l^{2}}{C_{0}(r)r^{2}sin^{2}\theta}.
\end{equation}

The $\theta -$ equation in the affine parameter can be obtained by differentiating this equation. Our strategy is to consider $\dot{\theta}^{2}\approx 0,$ for $\theta\approx\pi/2$, after the derivation. The constraint condition, however, amounts to the relation between the constants of the motion given by

 \begin{equation}
\frac{E^{2}}{l^{2}}\approx\frac{(r-r_{1})^{2\gamma-1}(r-r_{\Delta})^{2\gamma-1}}{K^{4}r^{2+4\gamma}},
\end{equation}
which is meaningful for $\gamma>1/2$. We note that with the choice $\theta\approx\pi/2$, we have

 \begin{equation}
\Sigma\approx1-\frac{2m}{r}+\frac{m^{2}}{r^{2}}\approx\Delta+\frac{m^{2}p^{2}}{r^{2}}.
\end{equation}

The second order equation for $\theta$ can be expressed upon certain expansions in the form

\begin{equation}
(r-r_{1})^{2(1-\gamma)}(r-r_{\Delta})^{2(1-\gamma)}\left(1+(1-\gamma^{2})\frac{m^{2}p^{2}}{r^{2}\Delta} \right)\ddot{\theta}\approx \frac{l^{2}cos\theta}{K^{4}r^{2(2\gamma-1)}sin^{3}\theta}.
\end{equation}
For a physical coefficient of $\ddot{\theta}$, which is finite and does not diverge for $r>r_{\Delta}$, we must have $\gamma<1$. This, however, must be limited with the coefficient on the right hand side, suggesting that $\gamma$ must satisfy $\gamma>1/2$. This result cocincides also with the result that was deduced from section (A) for $\omega^{2}$. In conclusion, our geodesics with circular, null character can be considered in the equatorial plane $\theta=\pi/2$, under the restriction that $1/2<\gamma<1.$ The restricted, approximate equation for $\theta-$ geodesics (49) is satisfied in this restricted sense. In the next section, we shall consider such restricted geodesics  to find the lensing effect of the charged $ZV$ spacetime.

\subsection{The Particle Motion in the Equatorial Plane }
In this subsection, we will consider the motion of magnetically and electrically charged test particles in charged $ZV$ spacetime separately.
\subsubsection{The magnetic case}
The motion of a test particle with charge $Q$ and unit mass is described by the Lagrangian

\begin{equation}
\mathcal{L}=\frac{1}{2}g_{\mu\nu}\frac{dx^{\mu}}{d\tau}\frac{dx^{\nu}}{d\tau}+QA_{\mu}\frac{dx^{\mu}}{d\tau}
\end{equation}
where the derivations are with respect to the proper time $\tau$. We choose our $ZV$ source with the magnetic field from (24), $A_{\mu}=(0,0,0,C_{0}cos\theta)$, so that the Euler-Lagrange equations give

\begin{equation}
g_{00}\left(\frac{dt}{d\tau} \right)=E=const.
\end{equation}

\begin{equation}
g_{\phi\phi}\left(\frac{d\phi}{d\tau} \right)+QC_{0}cos\theta=l=const.
\end{equation}

whereas the $\theta-$ equation is automatically satisfied in the equatorial plane. Now, we make the choice $\theta=\pi/2$, which removes the coupling term of test particle with the metric so that the analysis will be identical to that of a chargeless (neutral) particle. From the time-like geodesics condition we have

\begin{equation}
1=\frac{E^{2}}{g_{00}}+g_{rr}\left(\frac{dr}{d\tau} \right)^{2}+\frac{l^{2}}{g_{\phi\phi}}
\end{equation}
which is equivalent to

\begin{equation}
E^{2}=\Delta^{\gamma^{2}-1}\Sigma^{1-\gamma^{2}}\left(\frac{dr}{d\tau} \right)^{2}+\frac{\Delta^{\gamma}}{K^{2}}\left(1+\frac{l^{2}\Delta^{\gamma-1}}{K^{2}r^{2}} \right).
\end{equation}

Since we have chosen $\theta=\pi/2$, we have

\begin{equation}
\Sigma=\left(1-\frac{m}{r} \right)^{2},
\end{equation}

\begin{equation}
\Delta=\left(1-\frac{m}{r} \right)^{2}-\frac{m^{2}p^{2}}{r^{2}}.
\end{equation}

For circular geodesics we have $\frac{dr}{d\tau}=0$, which identifies from (54), the condition $V=E$. The potential acting on a neutral particle (or charged particle by a magnetically charged $ZV$ star) is given by

\begin{equation}
V(r)=\frac{\Delta^{\gamma/2}}{K}\left(1+\frac{l^{2}\Delta^{\gamma-1}}{K^{2}r^{2}} \right)^{1/2}.
\end{equation}

This must also satisfy $\frac{dV}{dr}=0$, for the circular geodesics which determines the possible angular momenta $l$ and in turn the corresponding energy values can be found from $E=V$. Herein we are not interested in circular geodesics, rather we shall abide by the general structure of an effective potential defined by

\begin{equation}
2V_{eff}=E^{2}-1-\left(\frac{dr}{d\tau} \right)^{2}
\end{equation}
Upon substitution for $\left(\frac{dr}{d\tau} \right)^{2}$ from (54), we obtain the effective potential as follows

\begin{equation}
V_{eff}=\frac{1}{2}(E^{2}-1)+\frac{1}{2}\left(\frac{\Delta}{\Sigma} \right)^{1-\gamma^{2}}\left[-E^{2}+\frac{\Delta^{\gamma}}{K^{2}}\left(1+\frac{l^{2}\Delta^{\gamma-1}}{K^{2}r^{2}} \right)\right].
\end{equation}

As a limiting case we can check the case for $\gamma=1$ and $p=\frac{1}{2}.$ The effective potential becomes

\begin{equation}
V_{eff}=-\frac{m}{r}+\frac{l^{2}}{r^{2}}\left(1-\frac{2m}{r}\right).
\end{equation}

This corresponds to the effective potential experienced by a neutral particle in the equatorial plane of a particularly charged RN geometry. The radial dependence of the effective potential (59) for different values of $\gamma$, but specific values of $E$, $l$ and $m$ is shown in figure (2). Figure (2) illustrates also that at asymptotic infinity, effect of $\gamma$ becomes weaker. On the other hand, effect of $\gamma$ is more stronger in the near regions. Another notable consequence of the charge is that an increase in the value of charge causes an increase in the potential barrier.

\begin{figure}[H]
\centering
  \begin{tabular}{@{}cccc@{}}
    \includegraphics{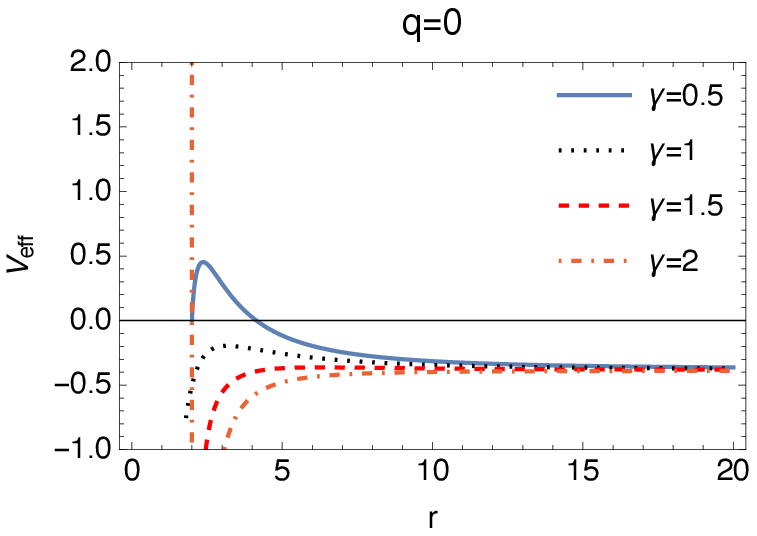} &
    \includegraphics{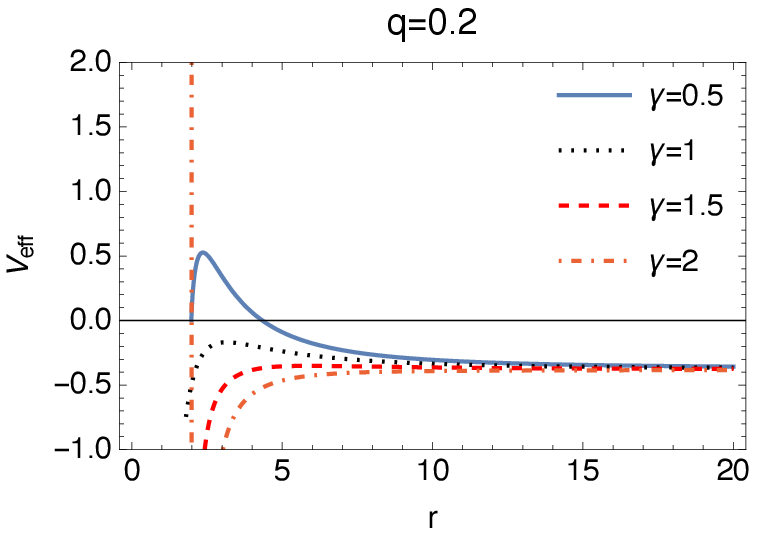} &
                                                         \\
    \includegraphics{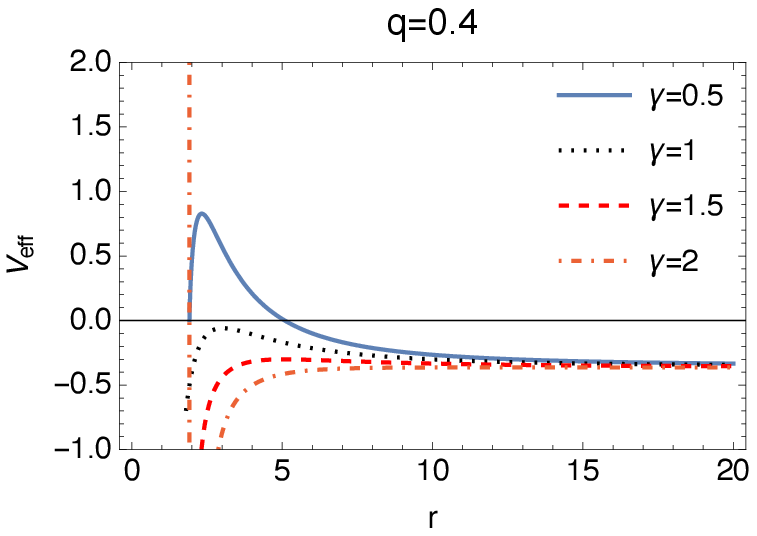} &
    \includegraphics{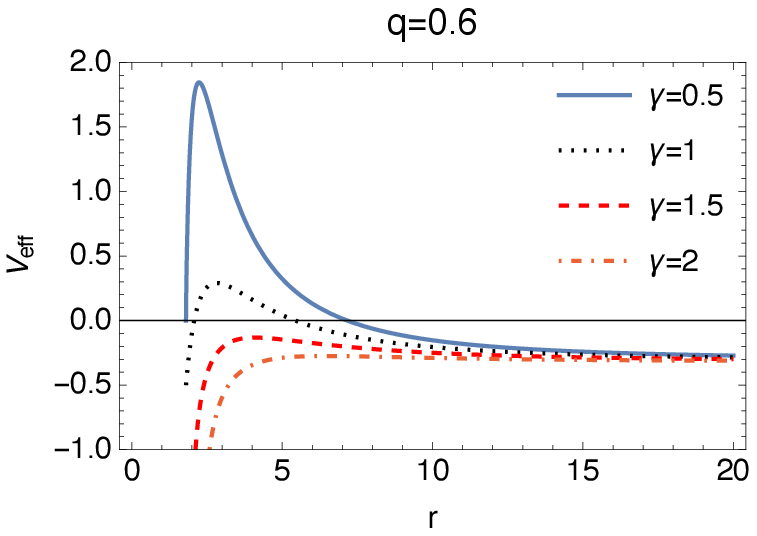} &
                                                           \\
     \multicolumn{2}{c}{\includegraphics{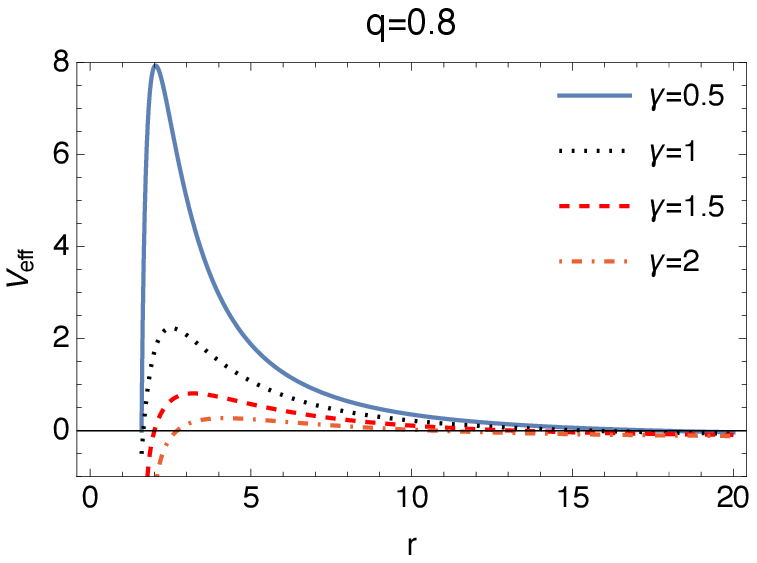}}
  \end{tabular}
  \caption{Plot for the effective potential (59) for specific parameters $m=1$, $E=1$ and $l=15.$ For comparison with the spherical case we added the special parameter $\gamma=1.$}
\end{figure}

\subsubsection{The electric case}

A test particle with electric charge $Q$ and unit mass for $\theta=\pi/2$ is described by the Lagrangian

\begin{equation}
\mathcal{L}=\frac{1}{2}g_{\mu\nu}\frac{dx^{\mu}}{d\tau}\frac{dx^{\nu}}{d\tau}+QA_{0}\frac{dt}{d\tau}
\end{equation}
where $A_{0}(r)$ is given in (30). From the Euler-Lagrange equations we obtain

\begin{equation}
\dot{t}=\frac{K^{2}}{\Delta^{\gamma}}(E-QA_{0}),
\end{equation}
and

\begin{equation}
\dot{\varphi}=\frac{l^{2}\Delta^{\gamma-1}}{r^{2}K^{2}},
\end{equation}
in which $E$ and $l$ are the integration constants for energy and angular momentum respectively. Note here also for $\theta=\pi/2$ the $\theta-$ equation is trivially satisfied. Following the similar steps as in the magnetic case, we obtain the effective potential

\begin{equation}
V_{eff}=\frac{1}{2}(E^{2}-1)+\frac{1}{2}\left(\frac{\Delta}{\Sigma} \right)^{1-\gamma^{2}}\left[-(E-QA_{0})^{2}+\frac{\Delta^{\gamma}}{K^{2}}\left(1+\frac{l^{2}\Delta^{\gamma-1}}{K^{2}r^{2}} \right)\right].
\end{equation}
where $\Sigma$ and $\Delta$ are same as in (55) and (56). Plots for the effective potential against radial distance $r$ are given in Fig.(3). As expected, effect of deformation parameter $\gamma$ at large distance becomes weaker, which is more effective in the near regions. As in the case of magnetic case, an increase in the charge value causes an increase in the potential barrier. It is also observed for $Q=0$, the $V_{eff}$ of electric case coincides with the $V_{eff}$ of the magnetic case.
\begin{figure}[H]
\centering
  \begin{tabular}{@{}cccc@{}}
    \includegraphics{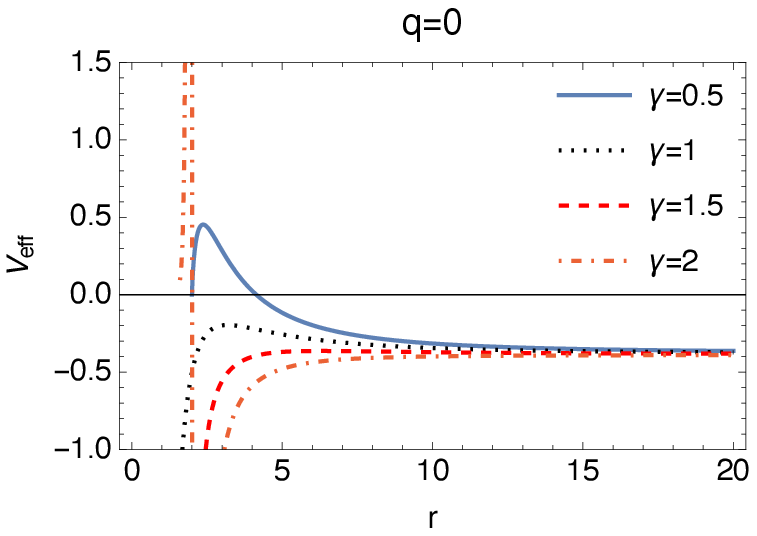} &
    \includegraphics{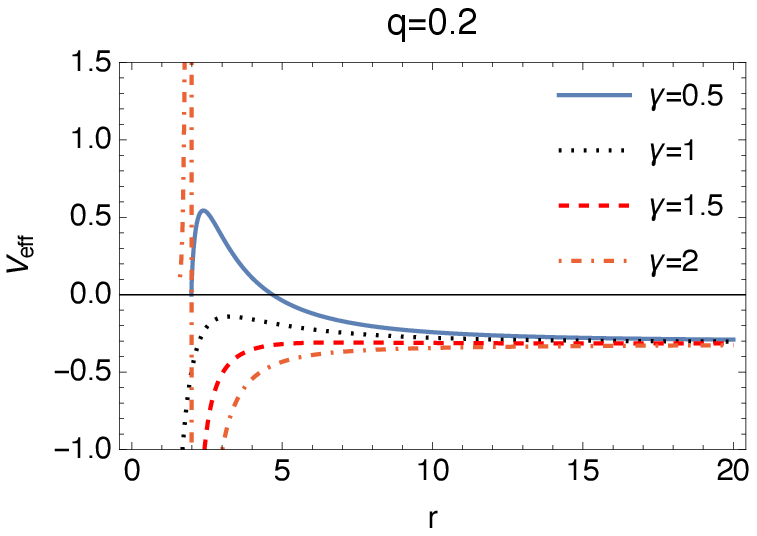} &
                                                         \\
    \includegraphics{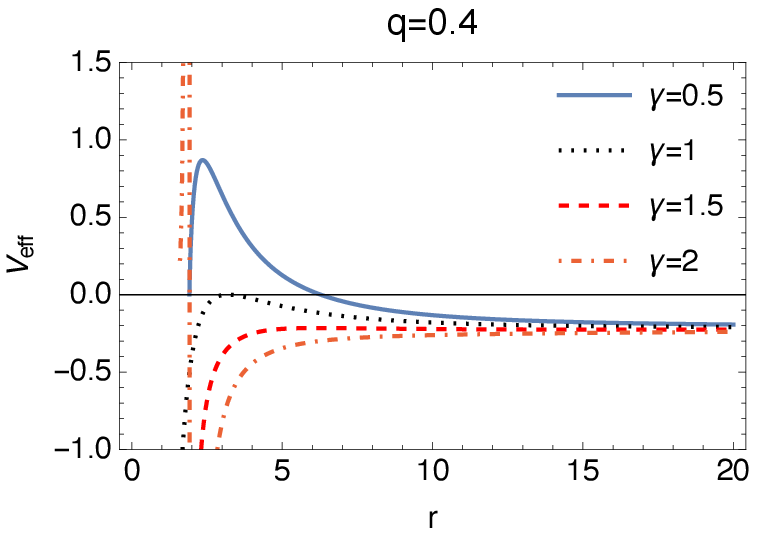} &
    \includegraphics{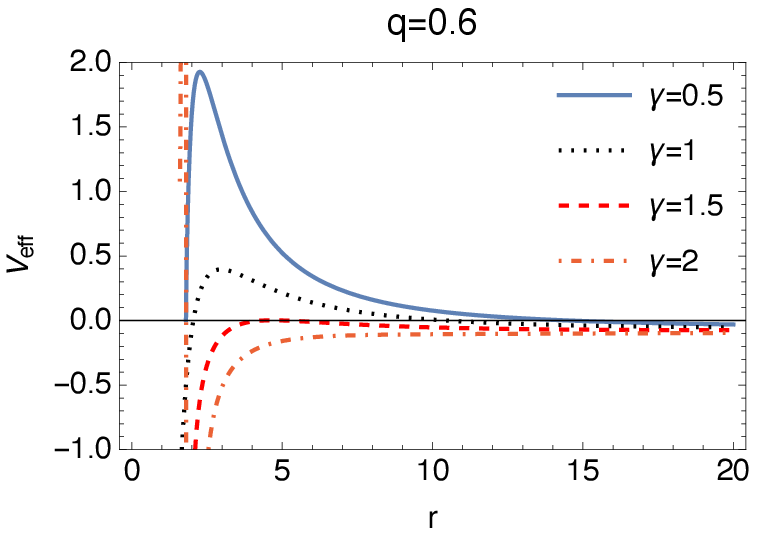} &
                                                           \\
     \multicolumn{2}{c}{\includegraphics{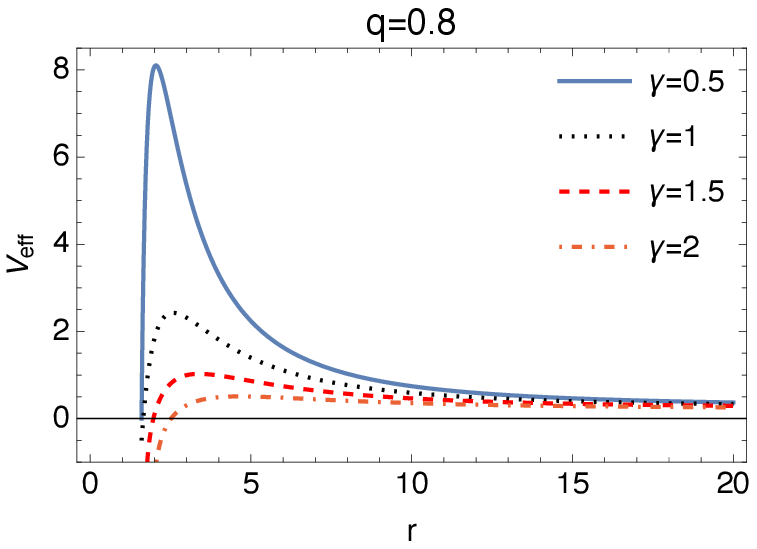}}
  \end{tabular}
  \caption{Radial dependence of the effective potential for a test particle with a charge $Q=1$, on the equatorial plane for different values of $\gamma$ and $q$. Plots are generated for specific values of $m=1$, $E=1$ and $l=15.$ }
\end{figure}

\section{Gravitational Lensing in the $ZV$ - Metrics}\label{sect5}

In this section, we will consider gravitational lensing for two different $ZV$ - metrics. In the first example, we will investigate the effect of the deformation parameter $\gamma$ on the gravitational lensing of the charged $ZV$ - metric. As a second example, stationary $ZV $ - metric will be investigated to clarify the role of spin on the bending angle together with the various values of the $\gamma$ parameter.

\subsection{Gravitational Lensing in Charged $ZV$ Spacetime}

In this subsection, we shall follow the method of Gibbons and Werner \cite{21} in order to calculate the gravitational lensing of light by the Gauss - Bonnet $(GB)$ theorem. See also \cite{22}. In the previous section  we considered the circular, null geodesics which can be projected to the $\theta=\pi/2$ plane. Note also that $\theta=\pi/2$ constitutes the symmetry plane of $ZV$  objects and one can project the geodesics into such  plane. Existence of circular geodesics allows also its perturbation into elliptical orbits which will be used in the present section. For this purpose, we introduce the optical metric of the spacetime $(18)$ as projected to the $\theta=\pi/2$ plane such that

\begin{equation}
dt^{2}=\bar{g}_{rr}dr^{2}+\bar{g}_{\varphi\varphi}d\varphi^{2},
\end{equation}
where

\begin{equation}
\bar{g}_{rr}=K^{4}\Delta^{\gamma^{2}-2\gamma-1}\Sigma^{1-\gamma^{2}},
\end{equation}

\begin{equation}
\bar{g}_{\varphi\varphi}=r^{2}K^{4}\Delta^{1-2\gamma}.
\end{equation}

The determinant of this metric has the square root

\begin{equation}
\sqrt{\bar{g}}=rK^{4}\Delta^{\frac{\gamma^{2}}{2}-2\gamma}\Sigma^{\frac{1-\gamma^{2}}{2}},
\end{equation}
and the Gaussian curvature

\begin{equation}
\mathcal{K}=-\frac{1}{\sqrt{\bar{g}}}\left[ \partial_{r}\left(\frac{1}{\sqrt{\bar{g}_{rr}}}\partial_{r} \sqrt{\bar{g}_{\varphi\varphi}}\right)+ \partial_{\varphi}\left(\frac{1}{\sqrt{\bar{g}_{\varphi\varphi}}}\partial_{\varphi} \sqrt{\bar{g}_{rr}}\right) \right],
\end{equation}
where the second term in the bracket vanishes for the present case. As explained in \cite{21}, $\mathcal{K}$ is the crucial expression for the application of the $GB$ theorem which states that the total deflection angle $\delta$ is given by

\begin{equation}
\int_{0}^{\pi}\int_{r_{g}}^{\infty}\mathcal{K}dS=-\delta,
\end{equation}
in which $dS=\sqrt{\bar{g}}drd\varphi$. Note that the lower limit $r_{g}$ in the integral is the angle dependent minimum distance from the source which is to be found from the null geodesics equation. In the present problem, we have

\begin{equation}
\int_{0}^{\pi}\int_{r_{g}}^{\infty}\mathcal{K}dS=-\int_{0}^{\pi}\int_{r_{g}}^{\infty}\partial_{r}\left[\Delta^{1-\frac{\gamma^{2}}{2}}\Sigma^{\frac{\gamma^{2}-1}{2}}\left(1+2r\frac{K^{\prime}}{K}+r\left(\frac{1}{2}-\gamma \right)\frac{\Delta^{\prime}}{\Delta} \right)   \right]drd\varphi.
\end{equation}

In order to calculate $\mathcal{K}$ we need expansions of the metric functions which are given as follow

\begin{equation}
\begin{aligned}
\Delta^{n}\simeq&1-\frac{2nm}{r}+\frac{nm^{2}}{r^{2}}\left[ q^{2}+2(n-1) \right]-\frac{2m^{3}}{3r^{3}}n(n-1)(3q^{2}+2)
\\
&+\frac{m^{4}}{r^{4}}n(n-1)\left[ \frac{1}{2}q^{4}+2q^{2}+\frac{2}{3}(n-2) \right]+... \; ,
\end{aligned}
\end{equation}

 \begin{equation}
\Sigma^{n}\simeq\Delta^{n}(q=1),
\end{equation}

 \begin{equation}
K\simeq 2p\left[ 1-\frac{m^{2}q^{2}}{2r^{2}}\gamma(\gamma-1)  \right]+... \; ,
\end{equation}

 \begin{equation}
\frac{\Delta^{\prime}}{\Delta} \simeq \frac{2m}{r^{2}}\left(1-\frac{mq^{2}}{r}  \right)\left[1+\frac{2m}{r}-\frac{m^{2}}{r^{2}}(q^{2}-4)  \right]+... \; ,
\end{equation}

 \begin{equation}
\frac{\Sigma^{\prime}}{\Sigma} \simeq \frac{2m}{r^{2}}\left(1+\frac{m}{r} +\frac{m^{2}}{r^{2}} \right)+... \; ,
\end{equation}

 \begin{equation}
\frac{K^{\prime}}{K}\simeq \frac{q^{2}m^{2}}{r^{3}}\gamma(\gamma-1)\left(1-\frac{m}{r}(\gamma-2)  \right)+... \; ,
\end{equation}
where "a prime" denotes $\frac{d}{dr}$. Note that the higher  order expansions are given in case further corrections are needed.

Next, in order to determine the lower limit $r_{g}$ we go to the null geodesics equation. We express the line element in the form (for $\theta=\pi/2$)

 \begin{equation}
ds^{2}=A(r)dt^{2}-B(r)dr^{2}-r^{2}C(r)d\varphi^{2},
\end{equation}
in which $A(r),B(r)$ and $C(r)$ are metric functions to be identified from (18). By introducing the geodesic Lagrangian parametrized by an affine paremeter $\lambda$

 \begin{equation}
\mathcal{L}=\frac{1}{2}A\dot{t}^{2}-\frac{1}{2}B\dot{r}^{2}-\frac{1}{2}r^{2}C\dot{\varphi}^{2},
\end{equation}
where, $\dot{}\equiv\frac{d}{d\lambda}$, yields the first integrals

 \begin{equation}
A\frac{dt}{d\lambda}=E=const. \; ,
\end{equation}

 \begin{equation}
Cr^{2}\frac{d\varphi}{d\lambda}=L=const. \; ,
\end{equation}
so that

 \begin{equation}
\frac{A}{Cr^{2}}\left(\frac{dt}{d\varphi}\right)=\frac{1}{b}=\frac{E}{L},
\end{equation}
where $b$ is the impact parameter. Shifting now to the new variable $u=1/r$, and solving $\left(\frac{dr}{d\lambda} \right)^{2}$ from $ds^{2}$ gives

 \begin{equation}
\left(\frac{du}{d\varphi}\right)^{2}=\frac{C}{B}\left(\frac{C}{Ab^{2}}-u^{2} \right),
\end{equation}
whose derivative with respect $\varphi$ yields the geodesics equation. By adopting the null geodesics equation into the present $ZV$ problem we obtain,

\begin{equation}
\begin{aligned}
\frac{d^{2}u}{d\varphi^{2}}+u=&3mu^{2}+\frac{16m(\gamma-1)p^{4}}{b^{2}}\left\{2-3mu\left[1-3\gamma+(1+\gamma)q^{2} \right]
\right.\\
&\left.+m^{2}u^{2}\left[ -3-3q^{2}(3+\gamma)(-1+2\gamma) \right]+\gamma(3+12\gamma+2p^{2}(-4+5\gamma))     \right\}.
\end{aligned}
\end{equation}

It is observed from this equation that for $\gamma=1$, the charge contribution comes at the order $1/b^{3}$ which is ignored. Upon solving the homogenous equation

 \begin{equation}
\frac{d^{2}u}{d\varphi^{2}}+u=0
\end{equation}
as $u=sin\varphi/b$, and plugging into the above equation we end up at the order $1/b^{2}$ with

 \begin{equation}
u=\frac{sin\varphi}{b}+\frac{32mp^{4}}{b^{2}}(\gamma-1)+\frac{m}{b^{2}}\left( 1+cos^{2}\varphi \right)
\end{equation}
which is to be identified as $1/r_{g}$.

Returning to Eqs.(69) and (70), using this limit for the $GB$ integral, we obtain the deflection angle (to the order $1/b^{2}$) as

 \begin{equation}
\delta=\frac{4m\gamma}{b}+\frac{\pi}{b^{2}}\left(m\gamma \right)^{2}\left[ 64p^{4}\left(1-\frac{1}{\gamma} \right)+\frac{4}{\gamma}-\frac{1}{4}-\frac{3}{4}q^{2} \right].
\end{equation}

In the limit $\gamma=1$, $q=0$ ($p=1$), one recovers the Schwarzschild deflection angle $\delta_{S}$

\begin{equation}
\delta_{S}=\frac{4m}{b}+\frac{15m^{2}\pi}{4b^{2}},
\end{equation}
and for the $RN$ limit, $\gamma=1$, $q\neq 0$ $\delta_{RN}$ as

\begin{equation}
\delta_{RN}=\frac{4m}{b}+\frac{15m^{2}\pi}{4b^{2}}-\frac{3Q^{2}\pi}{4b^{2}},
\end{equation}
where $Q=mq$. From (87), it is observed that with distortion parameter $\gamma$, one can define the new mass as $M=m\gamma$. Overall, the charge enters deflection formula at the second order correction to $\delta$. As an exceptional case note that for the specific charge, $q^{2}=5$, the $em$ contribution cancels with the deflection of the Schwarzschild term. Next, we consider in the sequel the example of  the stationary $ZV$ metric.

\subsection{Gravitational Lensing in the Stationary Uncharged $ZV$ Spacetime}

As a final example we consider the stationary solution for the $ZV$ metric without charge \cite{18}. For this purpose we use the metric from \cite{27}, where we accord the sign convention to ours

\begin{equation}
\begin{aligned}
ds^{2}=&e^{2\psi}dt^{2}-\frac{e^{2\lambda-2\psi}\Sigma}{\Delta}dr^{2}-e^{2\lambda-2\psi}\Sigma r^{2}d\theta^{2}\\
&-\left(e^{-2\psi}\Delta r^{2}sin^{2}\theta-\omega^{2}e^{2\psi} \right)d\varphi^{2}-2\omega e^{2\psi}dtd\varphi.
\end{aligned}
\end{equation}
where

\begin{equation}
\begin{aligned}
&e^{-2\psi}=\frac{1}{2}\left[(1-p_{0})\Delta^{\gamma}+(1+p_{0})\Delta^{-\gamma} \right]\\
&e^{2\lambda}=\left(\frac{\Delta}{\Sigma} \right)^{\gamma^{2}}\\
&\omega=-2m\gamma q_{0}cos\theta
\end{aligned}
\end{equation}
with $p_{0}^{2}+q_{0}^{2}=1$ and

\begin{equation}
\begin{aligned}
&\Delta=1-\frac{2m}{r}\\
&\Sigma=1-\frac{2m}{r}+\frac{m^{2}}{r^{2}}sin^{2}\theta.
\end{aligned}
\end{equation}

Note that $0<q_{0}<1$ is a NUT-like parameter that creates a cross term $\omega$ and since this term is proportional to $cos\theta$, it drops out for $\theta=\pi/2.$ Also in order to avoid confusion with our charged metric parameter of $p,$ we have labeled $p\rightarrow p_{0}$  $(q\rightarrow q_{0})$  of Ref.s \cite{18,27}. For $p_{0}=1$ this metric reduces to the $ZV$ metric. We note that in \cite{27}, the parameter $p_{0}$ is related to $q_{0}=\sqrt{1-p_{0}^{2}}$ dubbed as 'quasi-NUT', whereas in \cite{18} it was introduced as a differential 'spinning' parameter. This is due to the fact that unlike the NUT parameter $l$ ($0<l<\infty$), $p_{0}$ is strongly bounded, i.e. $p_{0}\leq1.$ From physical standpoint this may be considered applicable only for the largely extended astrophysical systems such as spiral galaxies

Once we fix $\theta=\pi/2$, the line element (90) leads to the optical metric

\begin{equation}
ds^{2}=e^{-4\psi}\left(\frac{e^{2\lambda}\Sigma}{\Delta}dr^{2}+\Delta r^{2}d\varphi^{2} \right)
\end{equation}
or

\begin{equation}
dt^{2}=\bar{g}_{rr}dr^{2}+\bar{g}_{\varphi\varphi}d\varphi^{2}.
\end{equation}

As in the charged $ZV$ case above, the deflection angle from the $GB$ theorem reduces to the expression

\begin{equation}
\delta=\int_{0}^{\pi}\int_{r_{g}}^{\infty}\left[ \partial_{r}\left(\frac{1}{\sqrt{\bar{g}_{rr}}}\partial_{r} \sqrt{\bar{g}_{\varphi\varphi}}\right)\right]drd\varphi.
\end{equation}

We note that such a reduction is possible due to the fact that although the optical metric (93) is not asymptotically flat (AF), for $\theta=\pi/2$ it becomes AF.

In order to determine the lower integration limit $r_{g}$, we use the null geodesics equation from (83) with the variable $u=1/r.$ With the impact parameter $b,$ we obtain after differentiating once more

\begin{equation}
\begin{aligned}
\frac{d^{2}u}{d\varphi^{2}}+u=&3mu^{2}+\frac{m^{2}}{b^{2}}\left[3+\gamma(5\gamma+4p_{0}(p_{0}\gamma-3)) \right]u\\
&+ \frac{2m}{b^{2}}(p_{0}\gamma-1)+\mathcal{O}\left(\frac{1}{b^{3}} \right).
\end{aligned}
\end{equation}
up to the order  $\sim \frac{1}{b^{2}}$, by using the homogenous solution as in the charged $ZV$ case, we obtain the lower limit of the $r-$ integral as

\begin{equation}
\frac{1}{r_{g}}=\frac{sin\varphi}{b}+\frac{m}{b^{2}}\left(2p_{0}\gamma-sin^{2}\varphi \right).
\end{equation}

Upon integrating (95), putting the limits of $r$ and integrating once more for $\varphi$ we obtain the deflection angle

\begin{equation}
\delta=\frac{4m\gamma p_{0}}{b}+\frac{m^{2}p_{0}\gamma\pi}{b^{2}}(4p_{0}\gamma-1)+\frac{m^{2}\gamma\pi}{4b^{2}}(4p_{0}+7\gamma-8p_{0}^{2}\gamma)
\end{equation}
which is valid only for $\theta=\pi/2.$ We observe that in the limit $\gamma=1$ (spherical symmetry) and $p_{0}=1$ ($q_{0}=0,$ the zero NUT-like parameter) we obtain $\delta_{S}.$

\section{Applications in Astrophysics}\label{sect6}

In this section, we study the obtained bending angle $\delta$ to explore the effect of the deviation parameter $\gamma$, both for the charged and the stationary uncharged $ZV$ - metrics. In doing so, we use the stellar data of charged compact objects \cite{24}, which were considered also in our earlier studies within the context of nonlinear electrodynamics \cite{25,26}. Observational estimation of the mass and the radius of the compact objects considered in this study are tabulated in Table-1 \cite{24}. The bending angle $\delta$ is plotted against  $x=b/R_{Star}$, here $R_{Star}$ denotes the estimated radius of the related charged compact object.

In figure 4-7, we have investigated numerically the effect of the deviation parameter $\gamma$ on the deflection angle $\delta$, for compact objects under consideration with charge and without charge. Due to the restriction on the value of the parameter $1/2<\gamma<1,$ three graphs are generated for each compact object, namely, for $\gamma=0.6$, $\gamma=0.8$ and $\gamma=1.$ Here, the case for $\gamma=1$ is plotted intentionally to be able to compare when the compact objects deviates from spherical symmetry. The figures 4-7 shows that irrespective whether the compact object is charged or uncharged, the deflection angle $\delta$ decreases when the spherical symmetry of the compact object tends to become more prolate. It should be noted that the effect of charge is almost negligible in the case of perfectly spherical compact objects $(\gamma=1)$. As a result the two plots are overlapped.

\begin{table}[th]
\caption{The numerical values of the masses and radii of the
 compact stars \cite{24}. In the table below $M_{\odot }$ represents the mass of the sun. }
\label{table:nonlin}
\centering
\begin{tabular}{|l|c|c|r|}
\hline
Compact Stars & M & Radius (km) \\ \hline\hline
Vela X-1  & $1.77M_{\odot }$ & $9.56$ \\ \hline
SAXJ 1808.4-3658  & $0.9M_{\odot }$ & $7.95$  \\ \hline
Her X-1  & $0.85M_{\odot }$ & $8.10$  \\ \hline
4U 1538-52  & $0.87M_{\odot }$ & $7.86$ \\ \hline
\end{tabular}
\end{table}

\begin{figure}[H]
\centering
  \begin{tabular}{@{}cccc@{}}
    \includegraphics{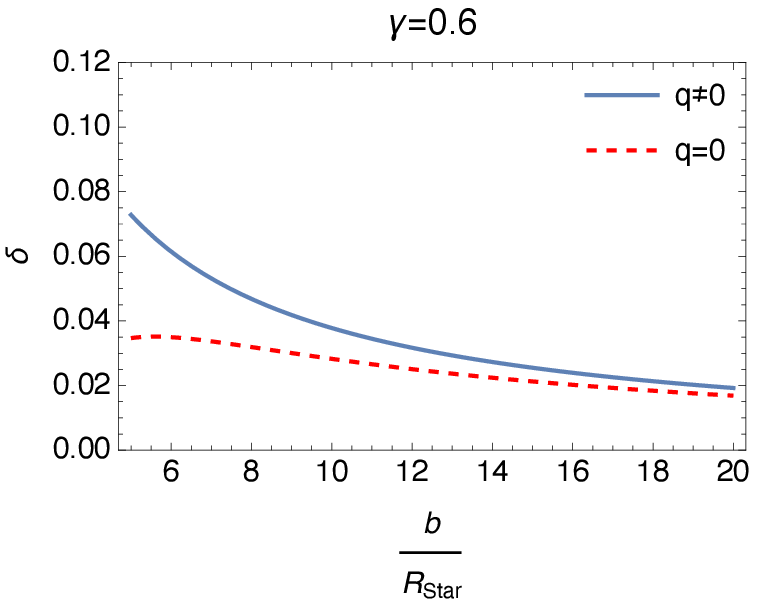} &
    \includegraphics{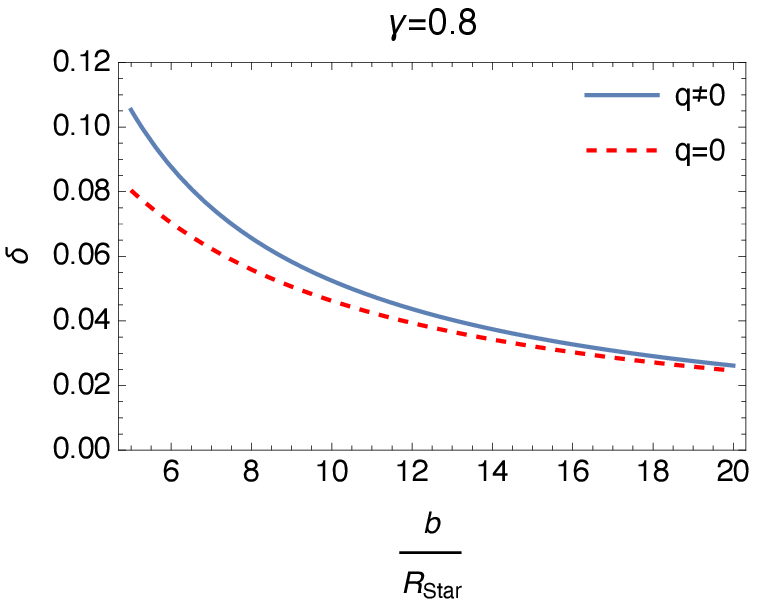} &
                                                                \\
    \multicolumn{2}{c}{\includegraphics{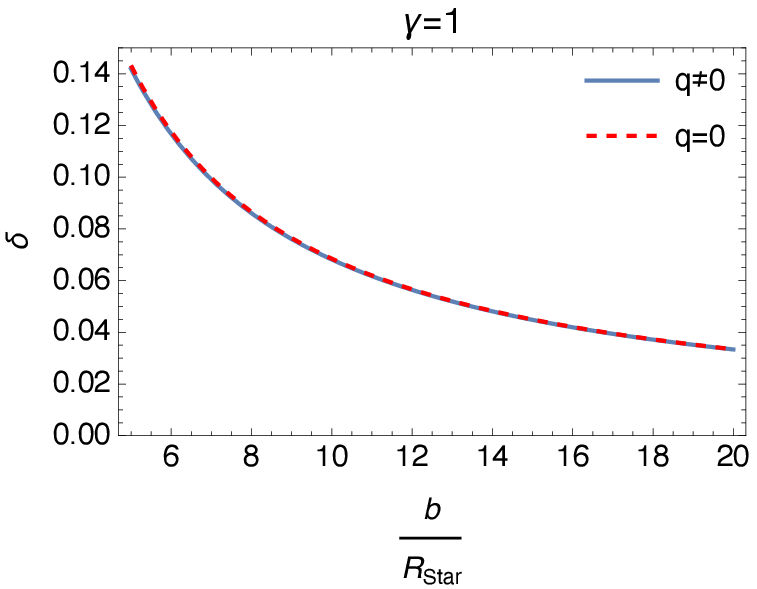}} &

  \end{tabular}
  \caption{The plot of deflection angle $\delta$ versus $x=b/R_{Star}$ for  $4U 1538-52$. Different curves correspond from left to right to the cases $\gamma=0.6$, $\gamma=0.8$ and $\gamma=1$. Note that in all plots we have chosen for the charge parameters $q=p=1/\sqrt{2}$. Solid line denotes charged, dashed  line denotes uncharged case.}
\end{figure}

\begin{figure}[H]
\centering
  \begin{tabular}{@{}cccc@{}}
    \includegraphics{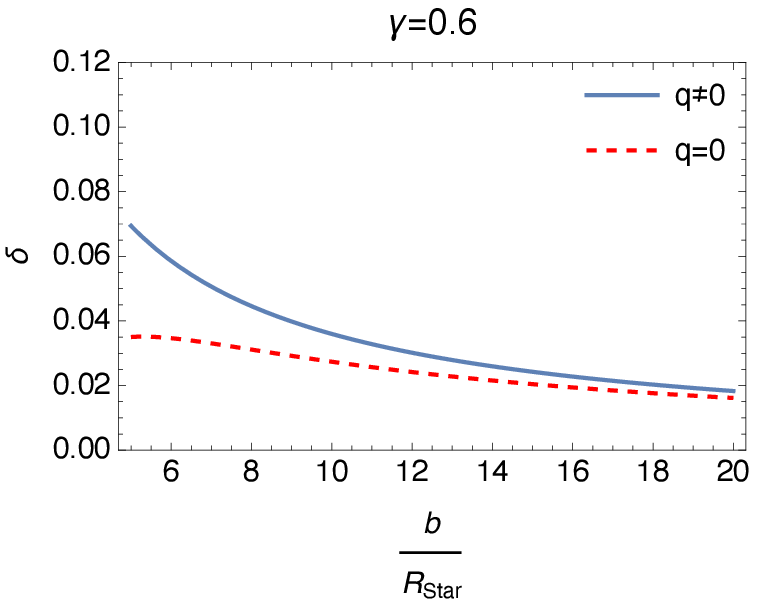} &
    \includegraphics{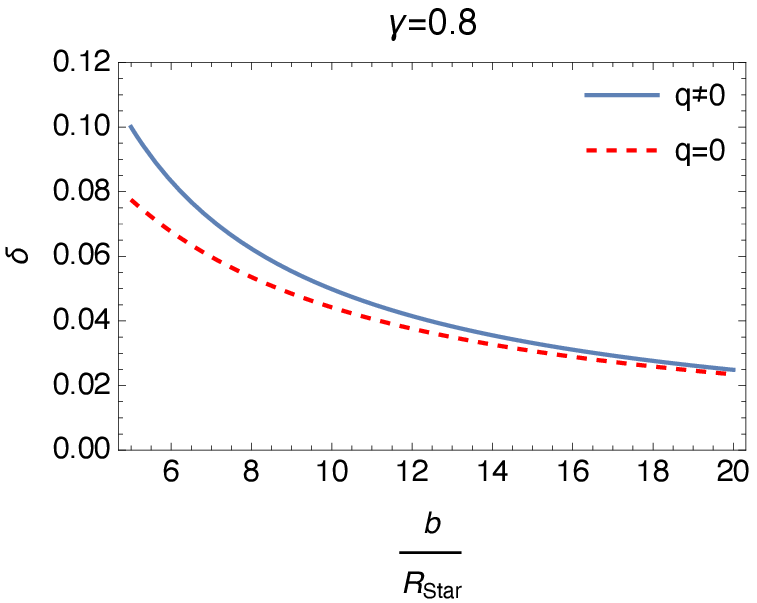} &
                                                                \\
    \multicolumn{2}{c}{\includegraphics{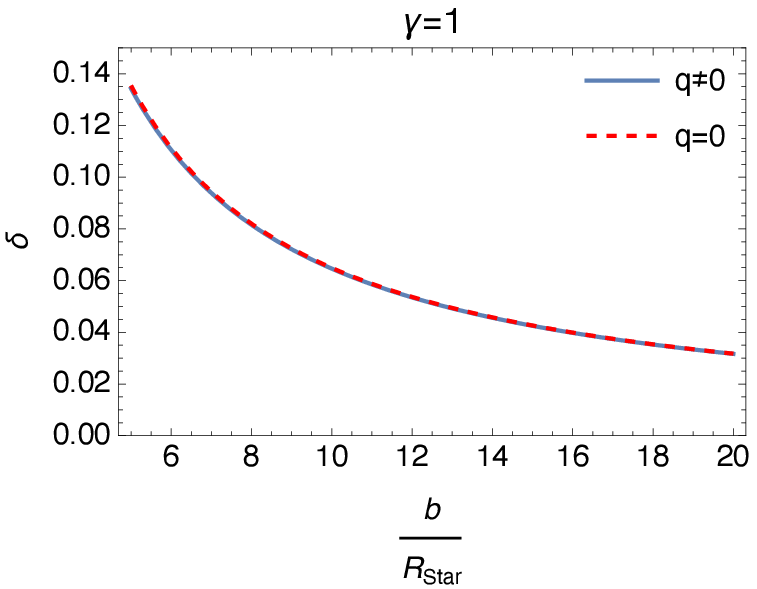}} &

  \end{tabular}
  \caption{$\delta$ versus $x$ plot for the $HerX-1$ star for the charged (solid line) / uncharged (dashed  line) cases.}
\end{figure}

\begin{figure}[H]
\centering
  \begin{tabular}{@{}cccc@{}}
    \includegraphics{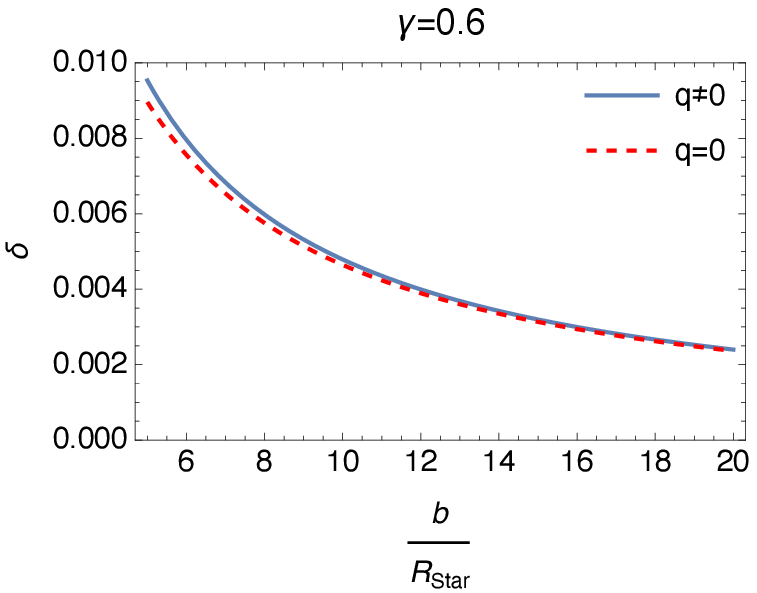} &
    \includegraphics{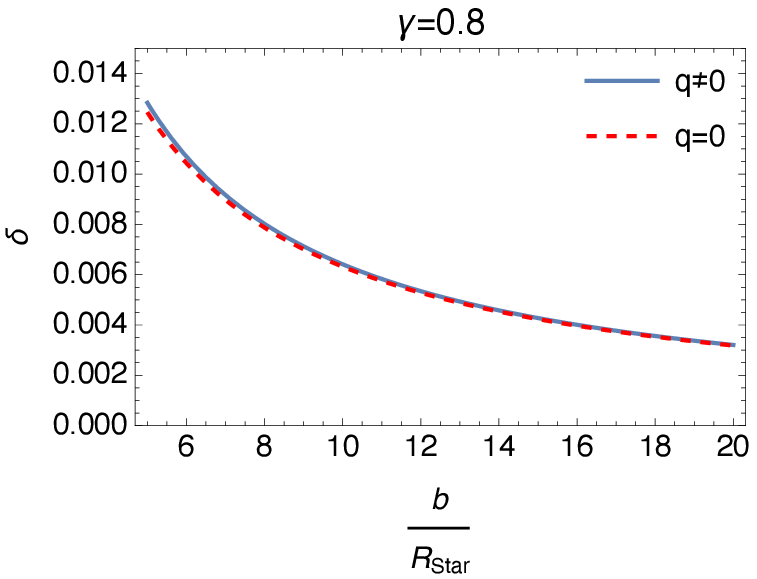} &
                                                                \\
    \multicolumn{2}{c}{\includegraphics{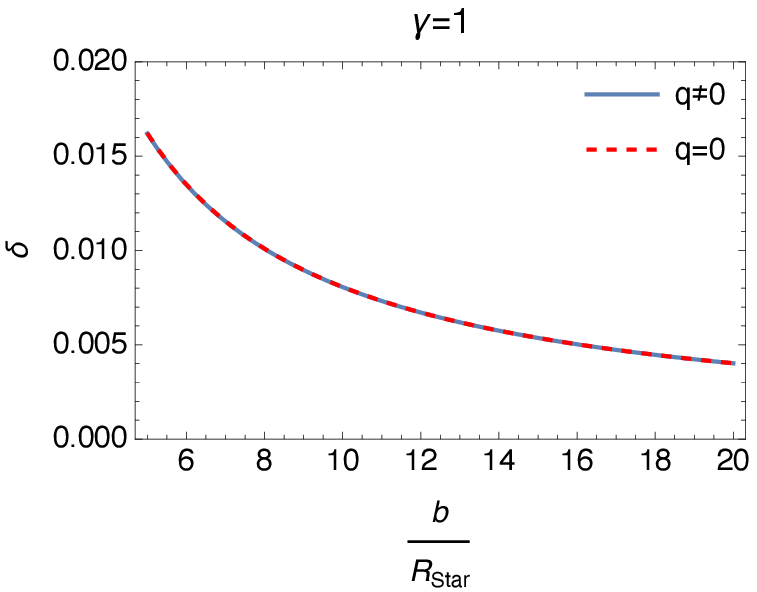}} &

  \end{tabular}
  \caption{Similar plot of $\delta$ versus $x$ for the $SAXJ1808.4-3658$ star.}
\end{figure}

\begin{figure}[H]
\centering
  \begin{tabular}{@{}cccc@{}}
    \includegraphics{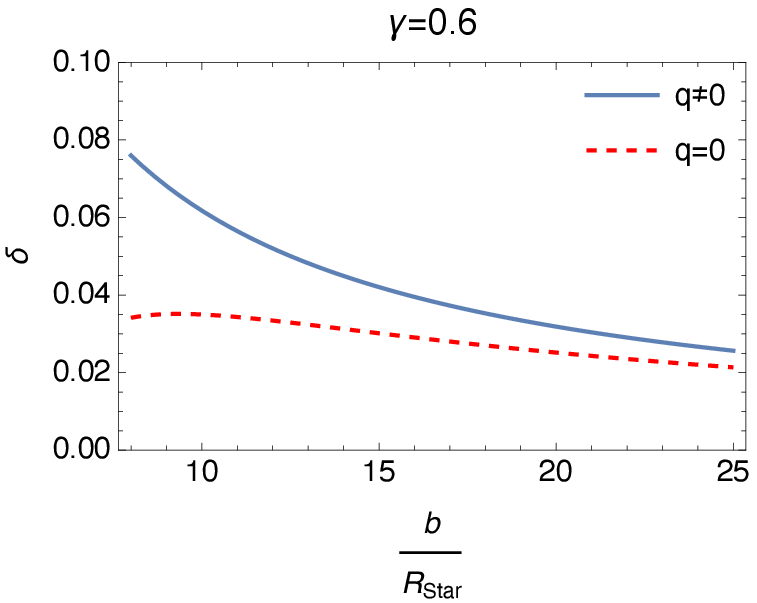} &
    \includegraphics{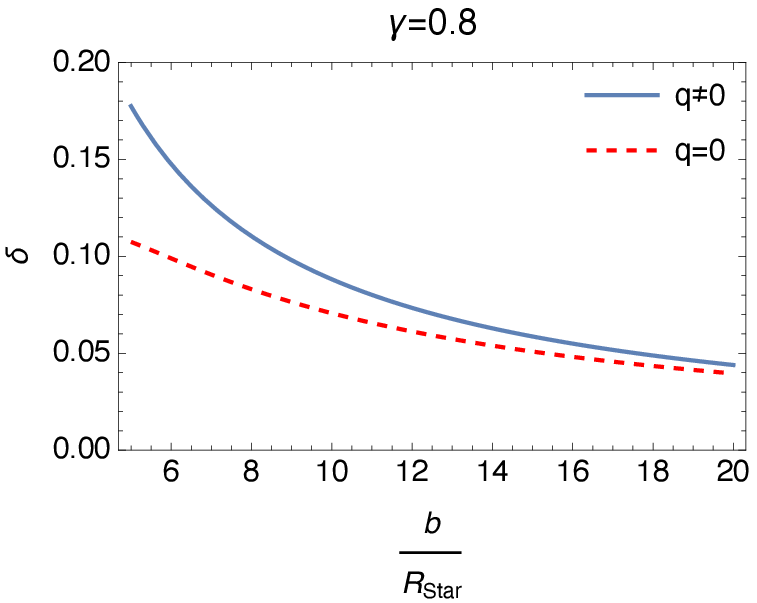} &
                                                                \\
    \multicolumn{2}{c}{\includegraphics{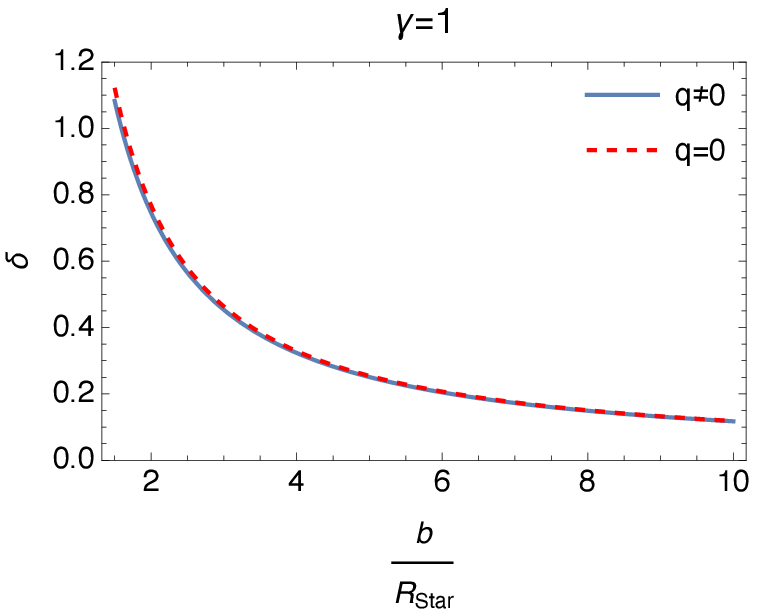}} &

  \end{tabular}
  \caption{$\delta$ versus $x$ plot for the  $VelaX-1$ star, both charged (solid line) and uncharged (dotted line) cases for comparison. }
\end{figure}

In figure 8, bending angle $\delta$ is plotted against $x$ for three different values of the deviation parameter $\gamma$. For each of the compact objects, bending angle $\delta$ increases as the distortion parameter increases. However, it is remained as an open problem, when the compact object tends to be more oblate $(\gamma>1)$, since, overall the system is not an integrable one.

\begin{figure}[H]
\centering
  \begin{tabular}{@{}cccc@{}}
    \includegraphics{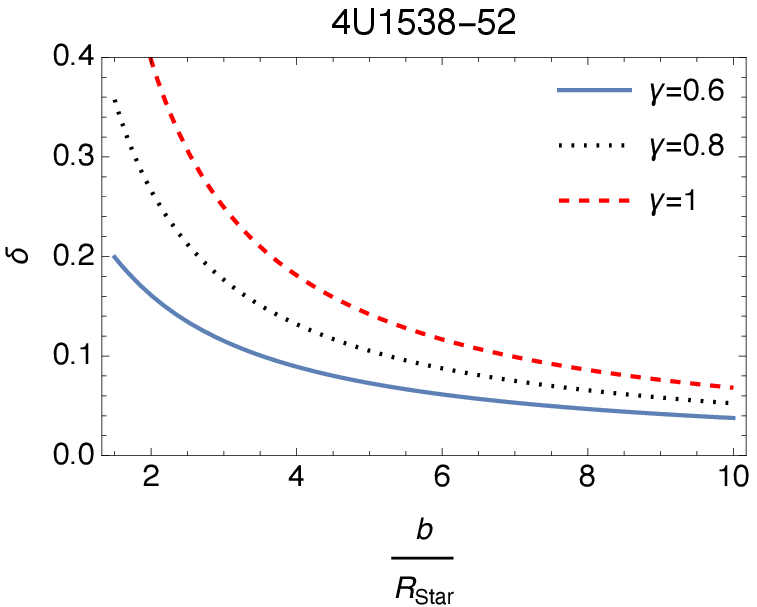} &
    \includegraphics{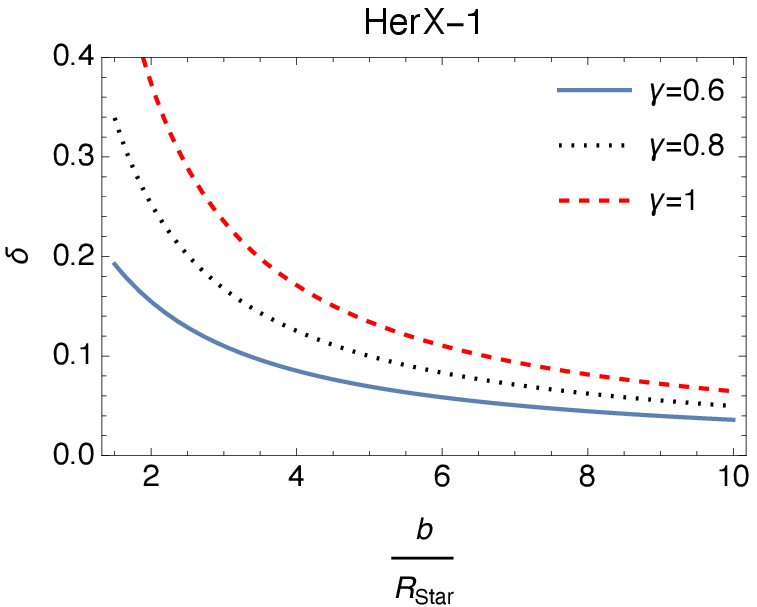} &
                                                         \\
    \includegraphics{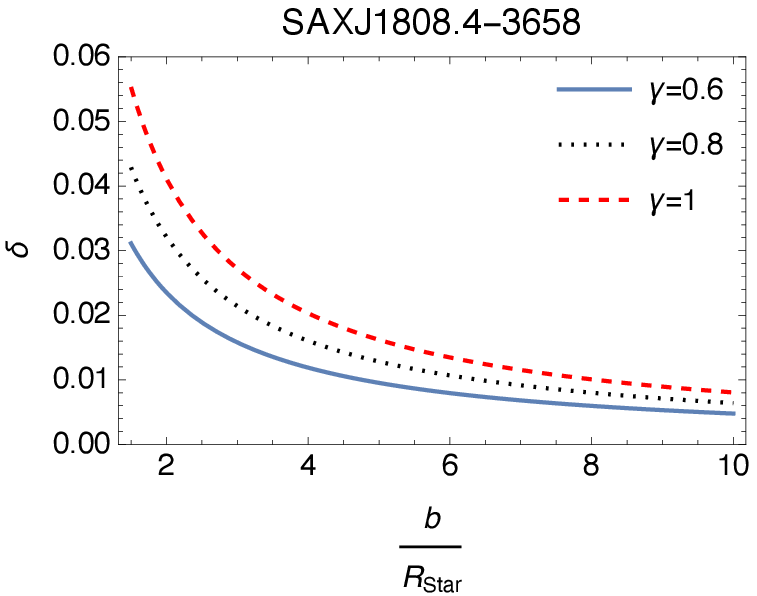} &
    \includegraphics{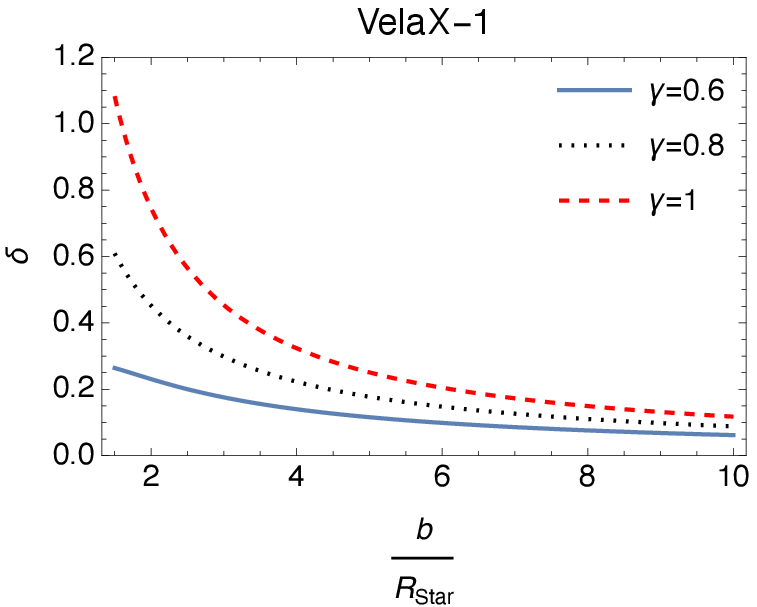} &
                                                            \\
    \end{tabular}
  \caption{Overlapping plots of $\delta$ versus $x$ for all types of stars in the Table 1, for $\gamma=0.6$ (solid line) , $\gamma=0.8$ (dotted line) and $\gamma=1$ (dashed line), for comparison. Plots are generated for the charged ZV case for a particular value of $p=q=\frac{1}{\sqrt{2}}$.}
\end{figure}

It is important to investigate also the gravitational lensing when  the non-spherical compact object is in the stationary state. In doing so, the calculated bending angle in Eq.(98) is studied numerically for the compact objects described in Table 1. \\
In figures 9-12, variation in the bending angle against idealized radial distance for different values of $\gamma$ parameter are plotted.  In these plots; uncharged stationary, uncharged and charged $ZV$ bending angles are shown together to clarify the 'spin' of the non-spherical compact  objects. Generated plots indicate that in the spherically symmetric case $\gamma=1$, charged and uncharged $ZV$ plots are overlapped and becomes hard to identify the difference. But, the gravitational lensing produced by the stationary uncharged $ZV$ metric can be identified easily for each compact star. From these plots we understand that the overall contribution of the 'spin' to the gravitational lensing of the compact object has a reverse effect so that reduces the bending angle. It is important to state that the plots for stationary uncharged case is generated for a particular value of $p_{0}=q_{0}=\frac{1}{\sqrt{2}}.$ Similarly, for the static charged case is plotted for $p=q=\frac{1}{\sqrt{2}}.$

\begin{figure}[H]
\centering
  \begin{tabular}{@{}cccc@{}}
    \includegraphics{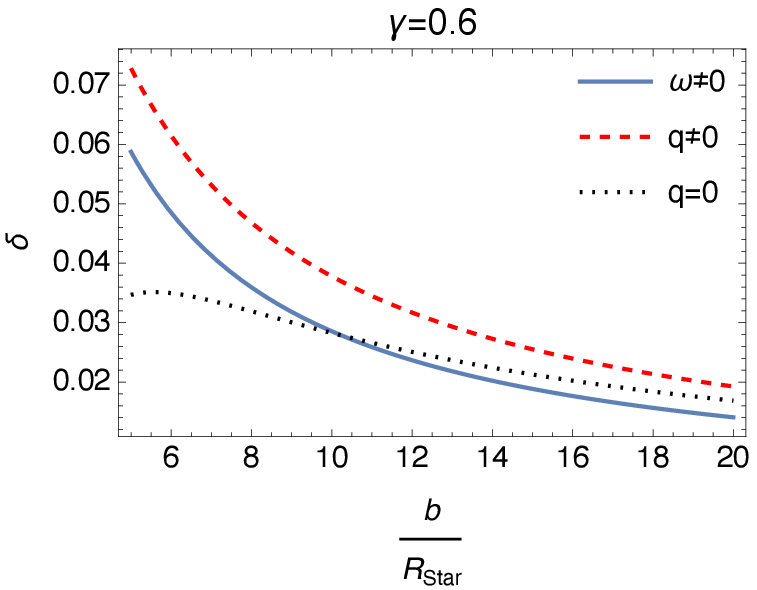} &
    \includegraphics{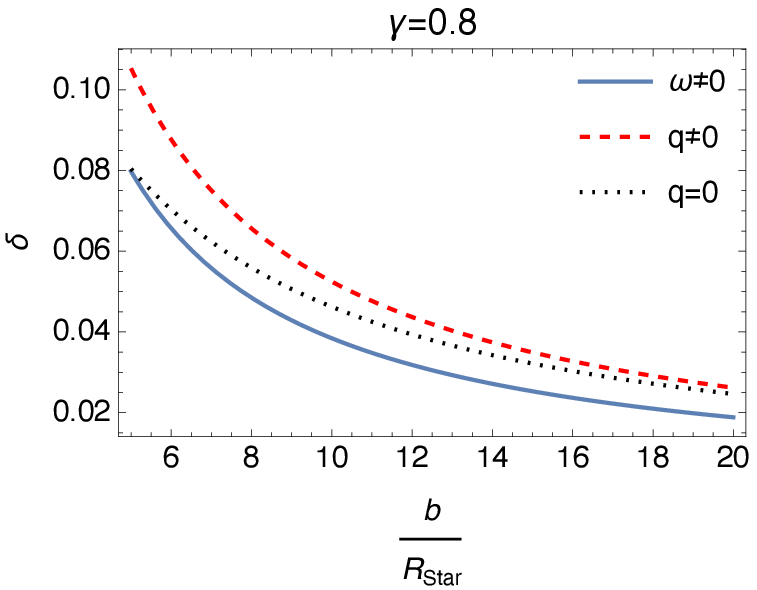} &
                                                                \\
    \multicolumn{2}{c}{\includegraphics{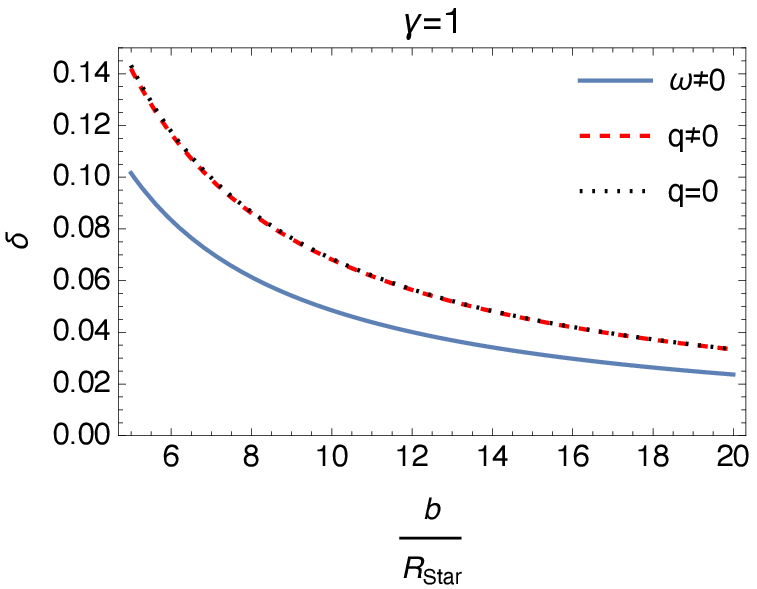}} &

  \end{tabular}
  \caption{Variation in the bending angle $\delta$ versus $x$ is plotted for the compact object 4U1538-52. The star is considered for the cases of uncharged stationary $(\omega\neq0)$, static charged $(q\neq0)$ and static uncharged $(q=0)$. Variation for each case are shown on the same graph for different values of deformation parameter $\gamma$. }
\end{figure}

\begin{figure}[H]
\centering
  \begin{tabular}{@{}cccc@{}}
    \includegraphics{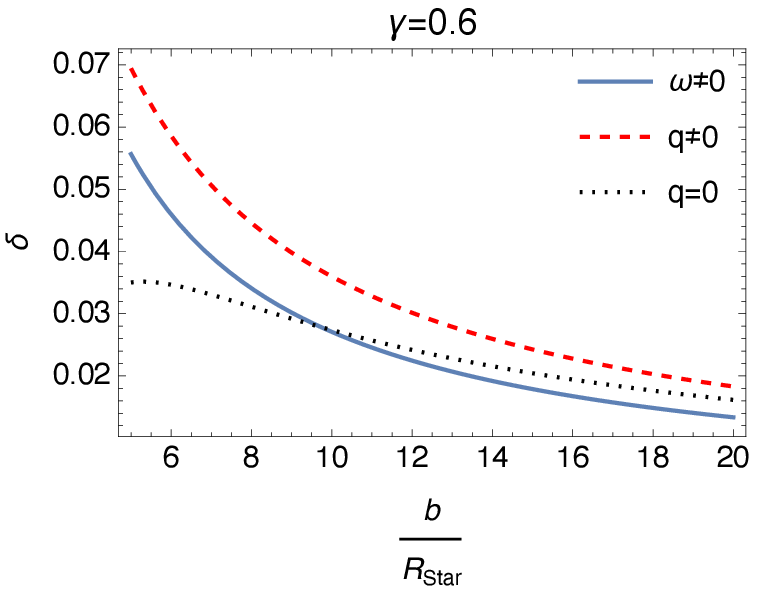} &
    \includegraphics{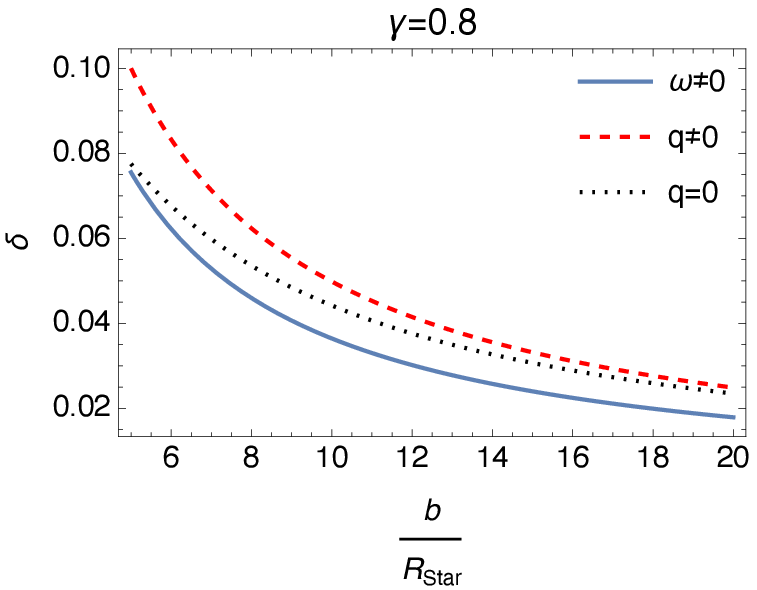} &
                                                                \\
    \multicolumn{2}{c}{\includegraphics{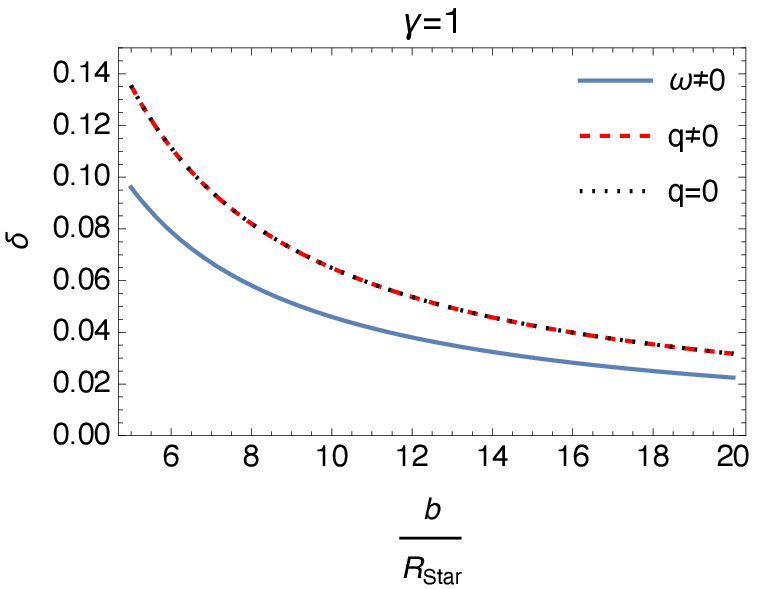}} &

  \end{tabular}
  \caption{Variation in the bending angle $\delta$ versus $x$ is plotted for the compact object HerX-1. The star is considered for the cases of uncharged stationary $(\omega\neq0)$, static charged $(q\neq0)$ and static uncharged $(q=0)$. Variation for each case are shown on the same graph for different values of deformation parameter $\gamma$.}
\end{figure}

\begin{figure}[H]
\centering
  \begin{tabular}{@{}cccc@{}}
    \includegraphics{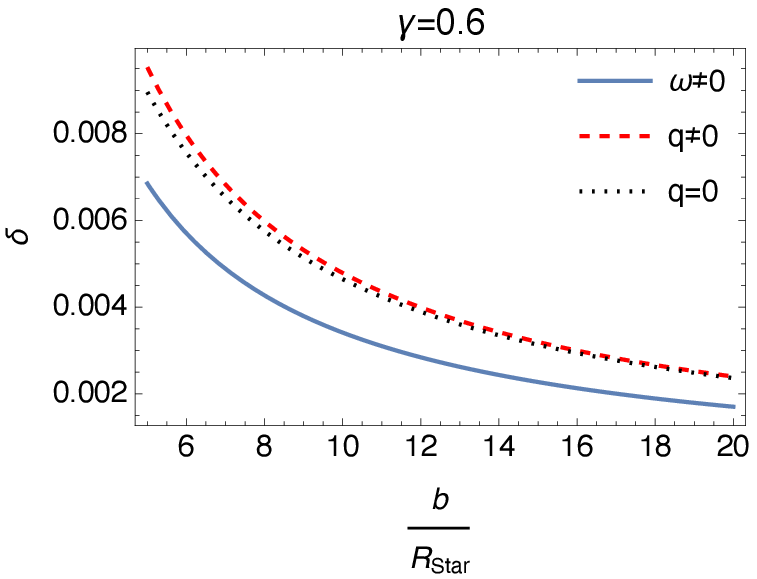} &
    \includegraphics{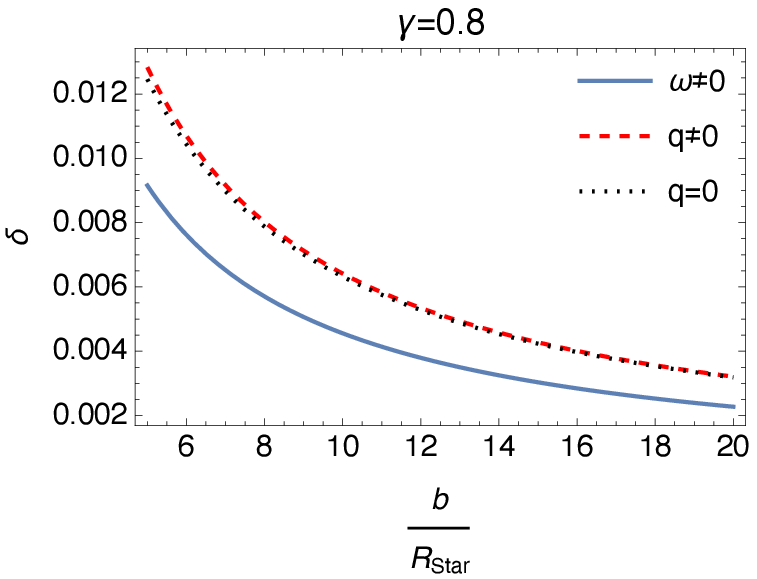} &
                                                                \\
    \multicolumn{2}{c}{\includegraphics{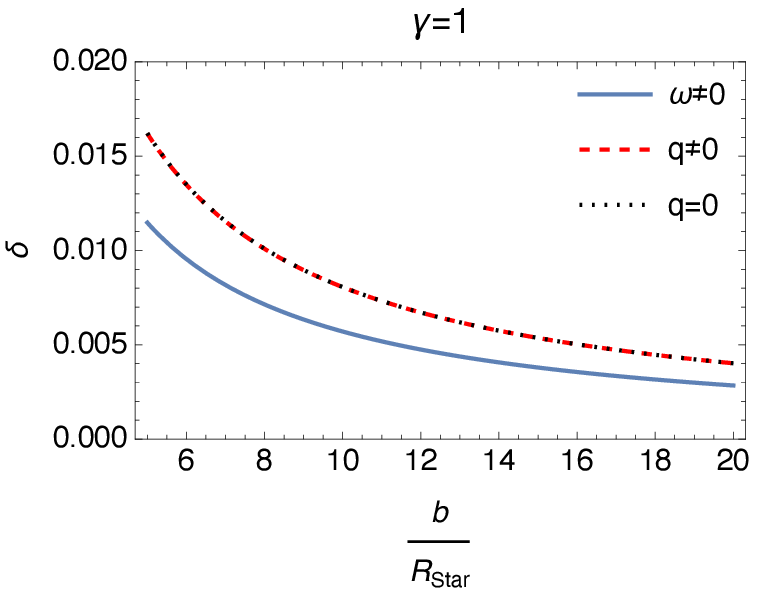}} &

  \end{tabular}
  \caption{Variation in the bending angle $\delta$ versus $x$ is plotted for the compact object SAXJ1808.4-3658. The star is considered for the cases of uncharged stationary $(\omega\neq0)$, static charged $(q\neq0)$ and static uncharged $(q=0)$. Note that bending angle for stationary state is less than the static cases for each particular $\gamma$ value.}
\end{figure}

\begin{figure}[H]
\centering
  \begin{tabular}{@{}cccc@{}}
    \includegraphics{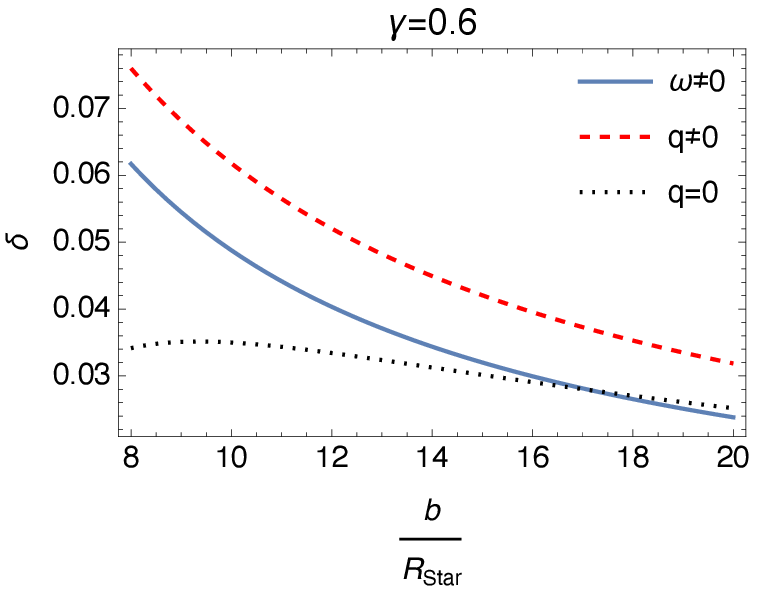} &
    \includegraphics{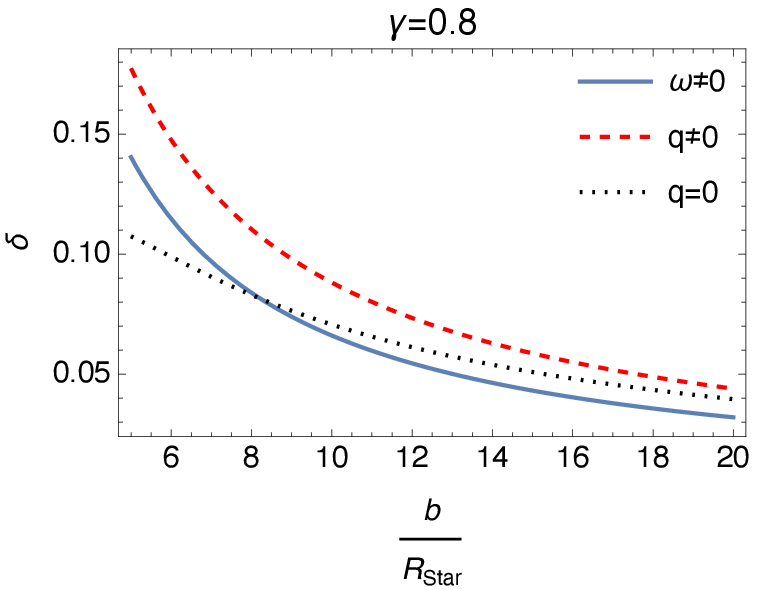} &
                                                                \\
    \multicolumn{2}{c}{\includegraphics{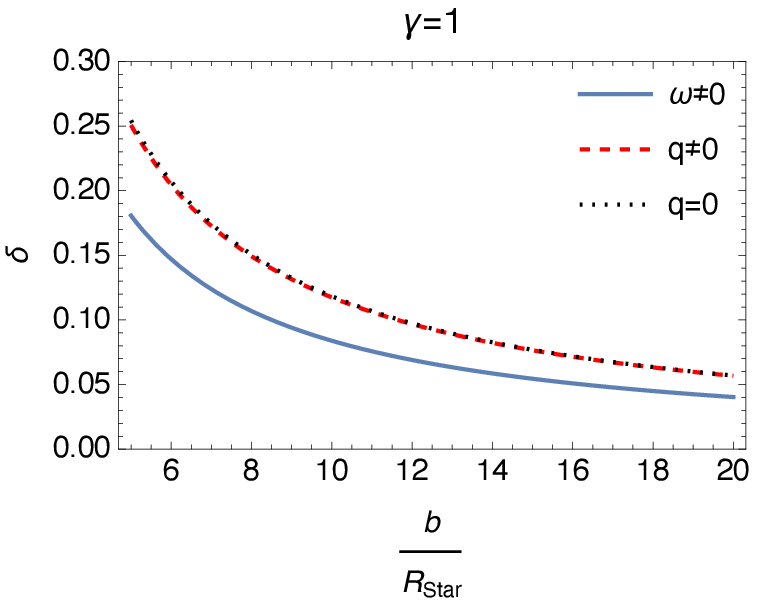}} &

  \end{tabular}
  \caption{Variation in the bending angle $\delta$ versus $x$ is plotted for the compact object VelaX-1. The star is considered for the cases of uncharged stationary $(\omega\neq0)$, static charged $(q\neq0)$ and static uncharged $(q=0)$ metrics. Variation for each case are shown on the same graph for different values of deformation parameter $\gamma$.}
\end{figure}

Figure 13 is generated to clarify the effect of the deformation parameter $\gamma$ on the gravitational lensing for the stationary state of uncharged $ZV$ metric.  According to the plots, the bending angle reduces as the deformation parameter $\gamma$ decreases.

\begin{figure}[H]
\centering
  \begin{tabular}{@{}cccc@{}}
    \includegraphics{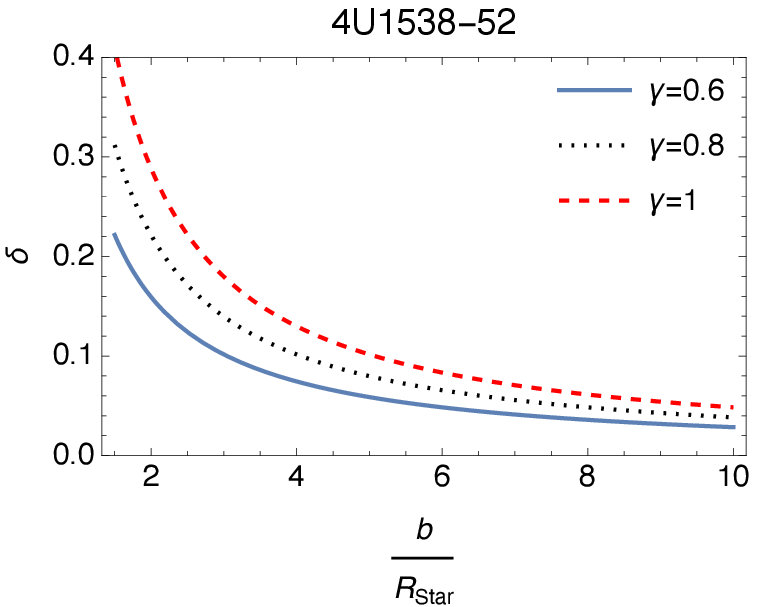} &
    \includegraphics{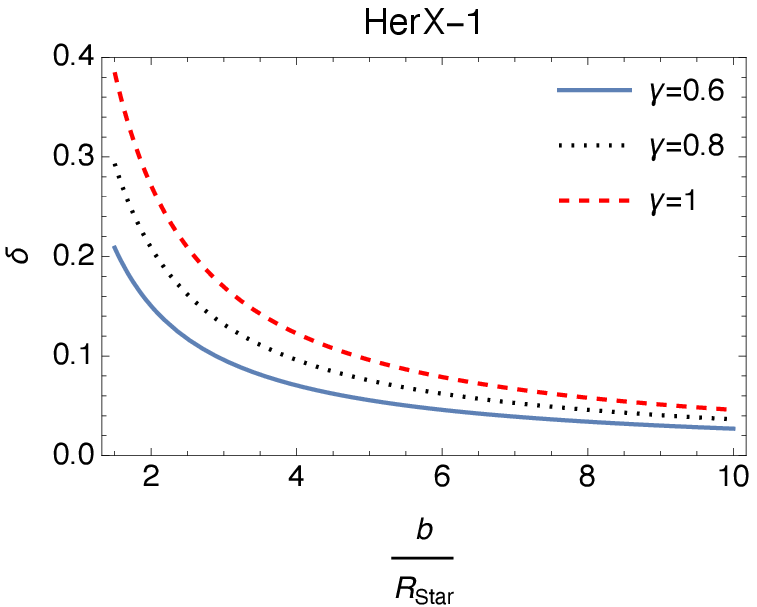} &
                                                         \\
    \includegraphics{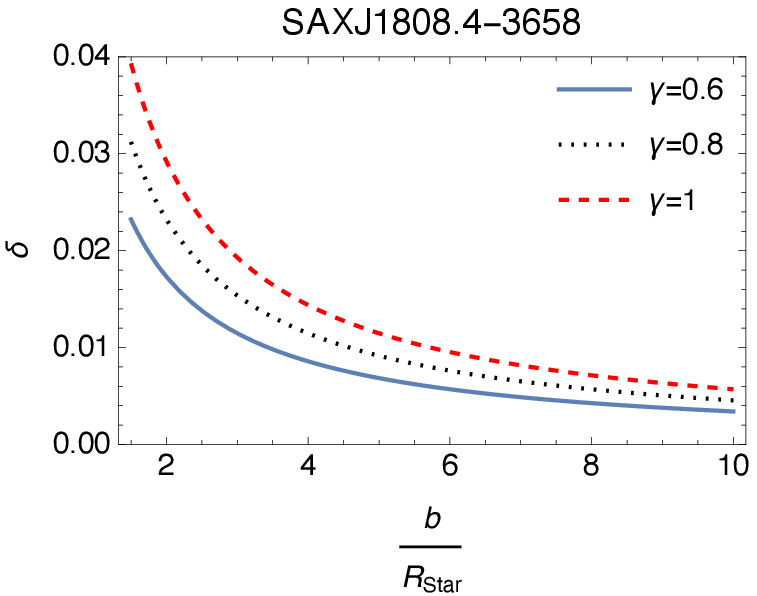} &
    \includegraphics{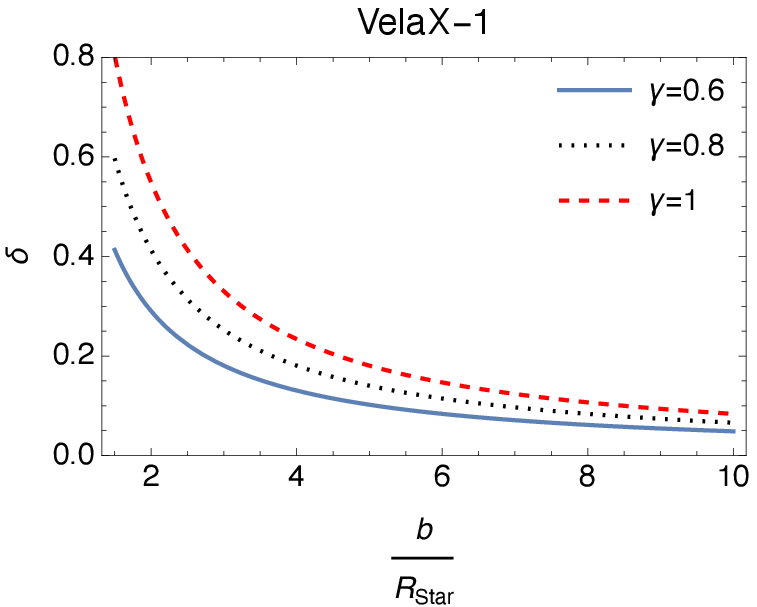} &
                                                            \\
    \end{tabular}
  \caption{Overlapping plots of the bending angle $\delta$ against normalized radial distance $x$ for compact stars given in Table 1, for different values of the parameter $\gamma$. Plots are generated for the stationary state of uncharged ZV metric.}
\end{figure}

Gravitational redshift is another remarkable observable parameter in astrophysics. We calculate gravitational redshift for the static non-spherical compact object with charged and uncharged cases. Our main concern will be on the effect of deformation parameter $\gamma.$

Our calculation will be based on the method developed for static cases in \cite{28,29}. According to this method the gravitational redshift formula is given by

\begin{equation}
z=\frac{\lambda_{o}-\lambda_{e}}{\lambda_{e}}=\frac{\lambda_{o}}{\lambda_{e}}-1=\frac{\omega_{o}}{\omega_{e}}-1
\end{equation}
where

\begin{equation}
\frac{\omega_{e}}{\omega_{o}}=\sqrt{g_{tt}}.
\end{equation}
Here, $\lambda_{o}$ and $\lambda_{e}$ denotes observed and emitted wavelengths, respectively. Similarly, $\omega_{o}$ and $\omega_{e}$ represents observed and emitted frequencies.

The explicit expression for the gravitational redshift is found as

\begin{equation}
z=\frac{(1+p)\left(1-\frac{m(1-p)}{r}   \right)^{\gamma}-(1-p)\left(1-\frac{m(1+p)}{r}   \right)^{\gamma}}{\left(1-\frac{2m}{r}+\frac{m^{2}q^{2}}{r^{2}} \right)^{\gamma/2}}-1
\end{equation}

The plots are generated for various values of the deformation parameter $\gamma$. The effect of charge is also displayed by comparing the gravitational redshift with the uncharged case.

\begin{figure}[H]
\centering
  \begin{tabular}{@{}cccc@{}}
    \includegraphics{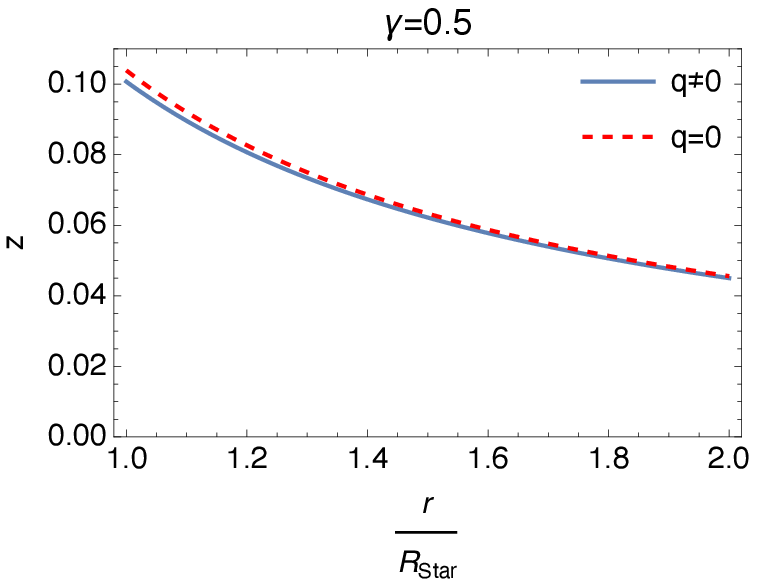} &
    \includegraphics{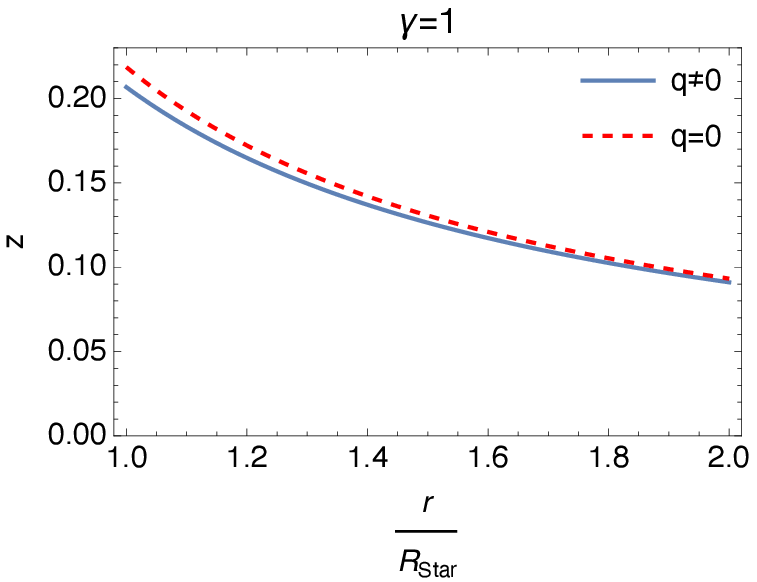} &
                                                                \\
    \multicolumn{2}{c}{\includegraphics{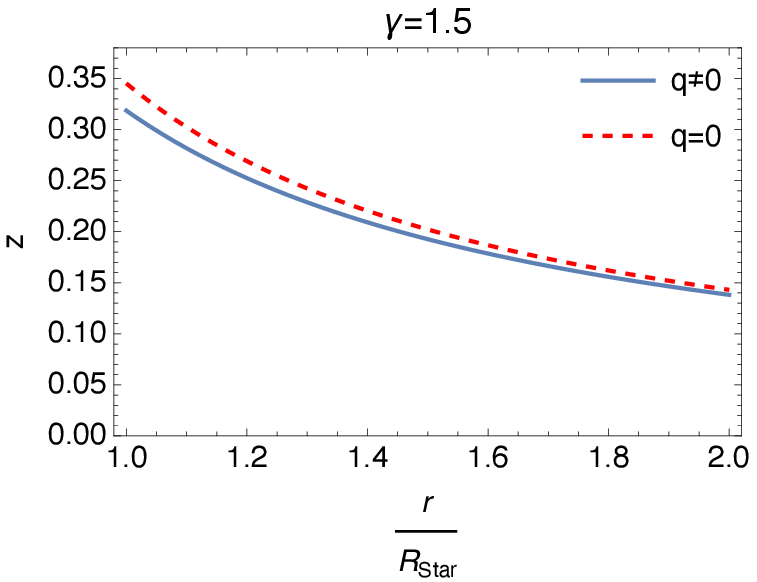}} &

  \end{tabular}
  \caption{The graphs show the variation in the redshift $z$ versus $\frac{r}{R_{Star}}$ for the compact object 4U1538-52. The plots are generated for static charged $(q\neq0)$ and static uncharged $(q=0)$ cases. Variation for each case are shown on the same graph for different values of deformation parameter $\gamma$.}
\end{figure}

\begin{figure}[H]
\centering
  \begin{tabular}{@{}cccc@{}}
    \includegraphics{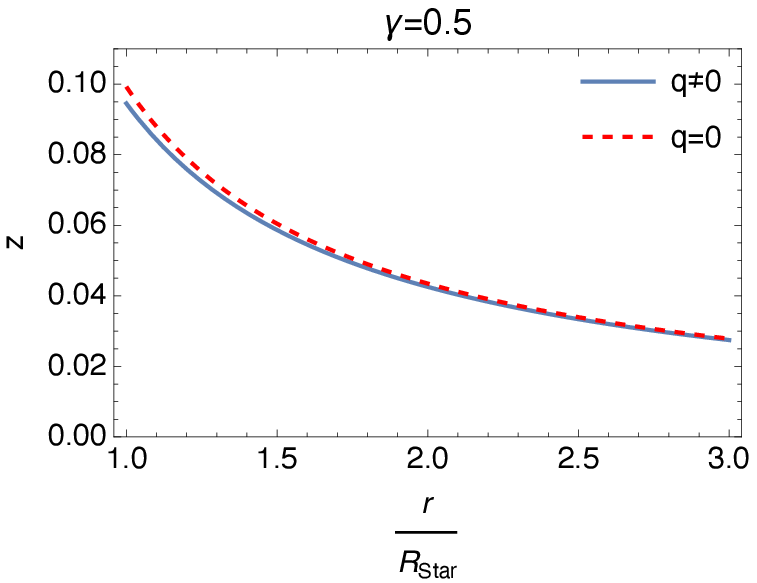} &
    \includegraphics{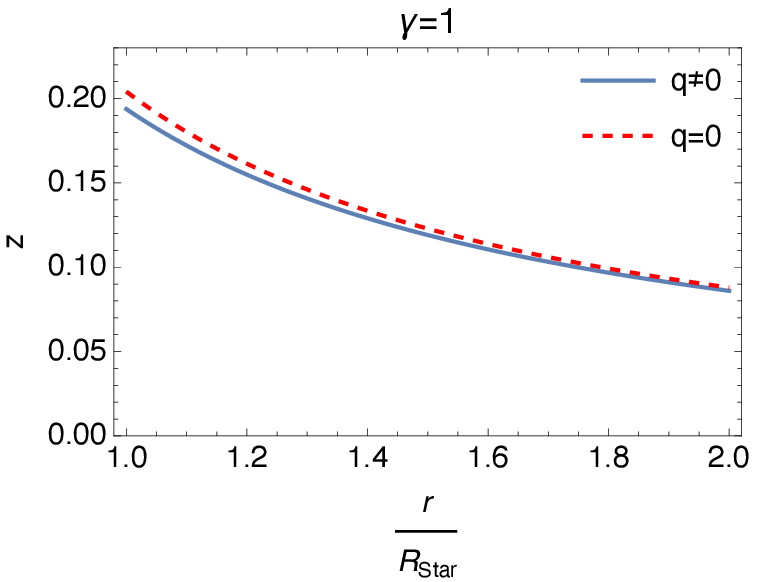} &
                                                                \\
    \multicolumn{2}{c}{\includegraphics{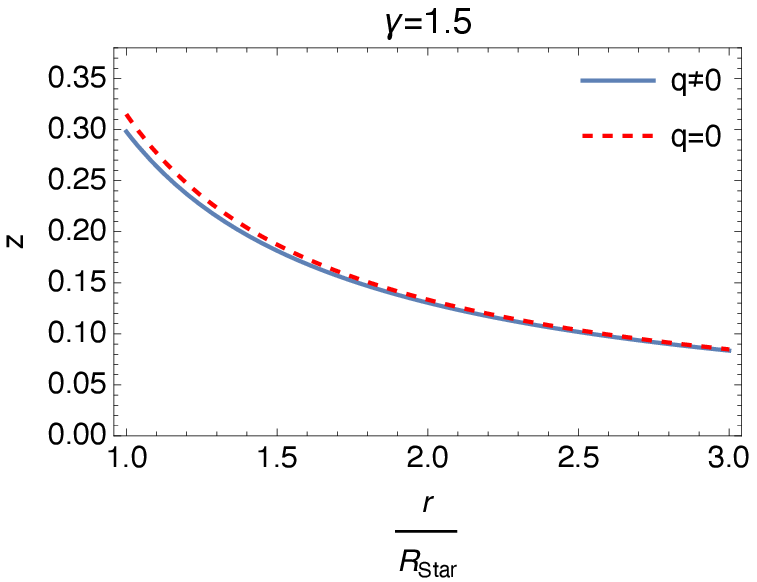}} &

  \end{tabular}
  \caption{The graphs show the variation in the redshift $z$ versus $\frac{r}{R_{Star}}$ for the compact object HerX-1. The plots are generated for static charged $(q\neq0)$  and static uncharged $(q=0)$  cases. Variation for each case are shown on the same graph for different values of deformation parameter $\gamma$.}
\end{figure}

\begin{figure}[H]
\centering
  \begin{tabular}{@{}cccc@{}}
    \includegraphics{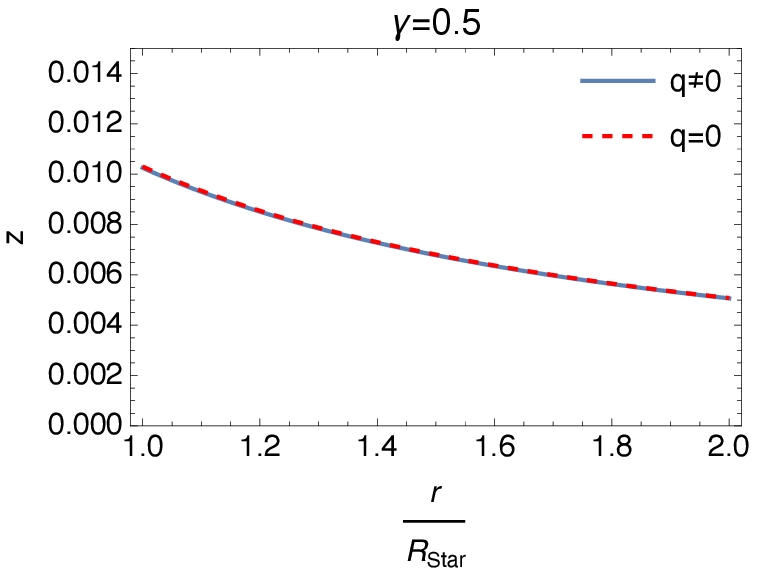} &
    \includegraphics{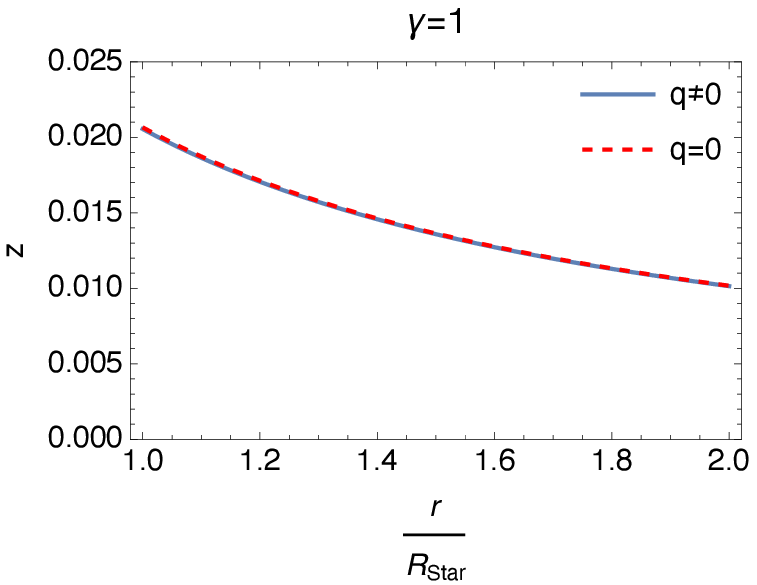} &
                                                                \\
    \multicolumn{2}{c}{\includegraphics{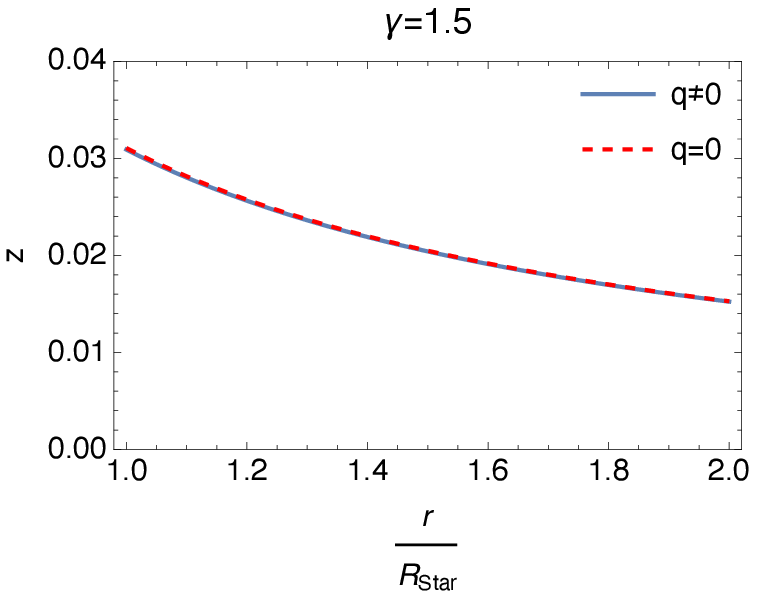}} &

  \end{tabular}
  \caption{The graphs show the variation in the redshift $z$ versus $\frac{r}{R_{Star}}$ for the compact object SAXJ1808.4-3658. The plots are generated for static charged $(q\neq0)$  and static uncharged $(q=0)$  cases. Variation for each case are shown on the same graph for different values of deformation parameter $\gamma$.}
\end{figure}

\begin{figure}[H]
\centering
  \begin{tabular}{@{}cccc@{}}
    \includegraphics{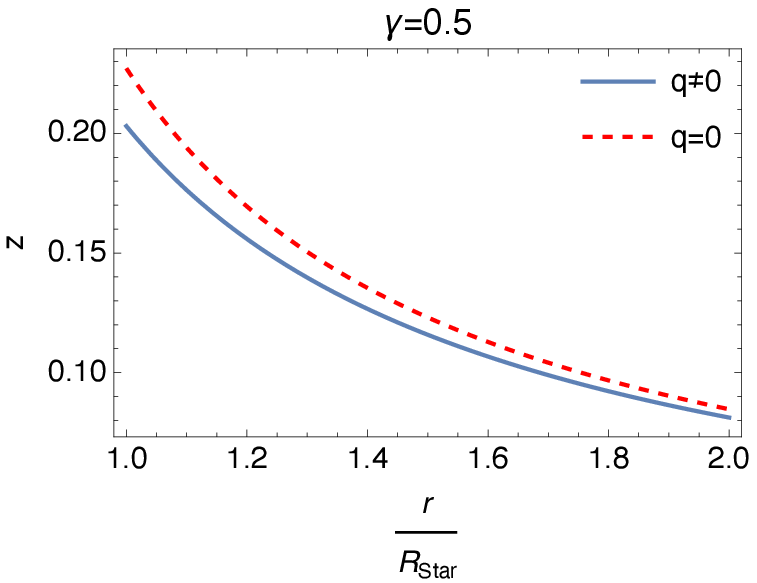} &
    \includegraphics{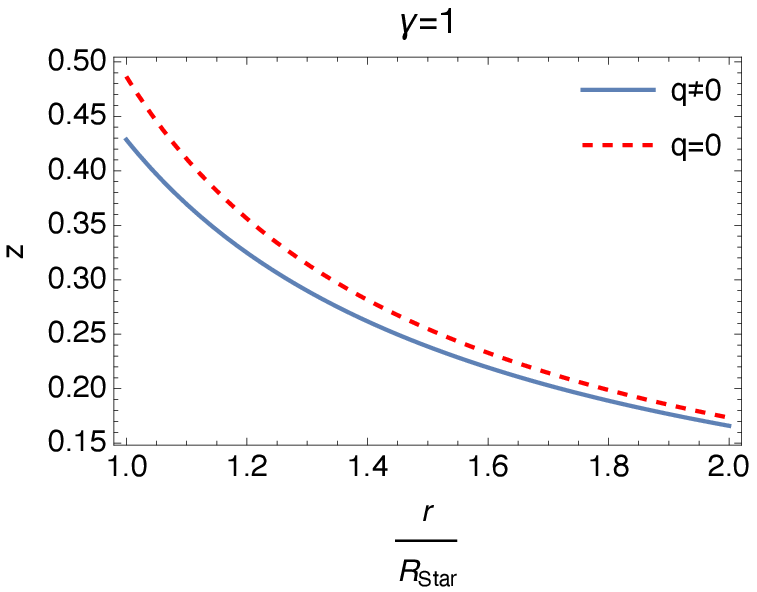} &
                                                                \\
    \multicolumn{2}{c}{\includegraphics{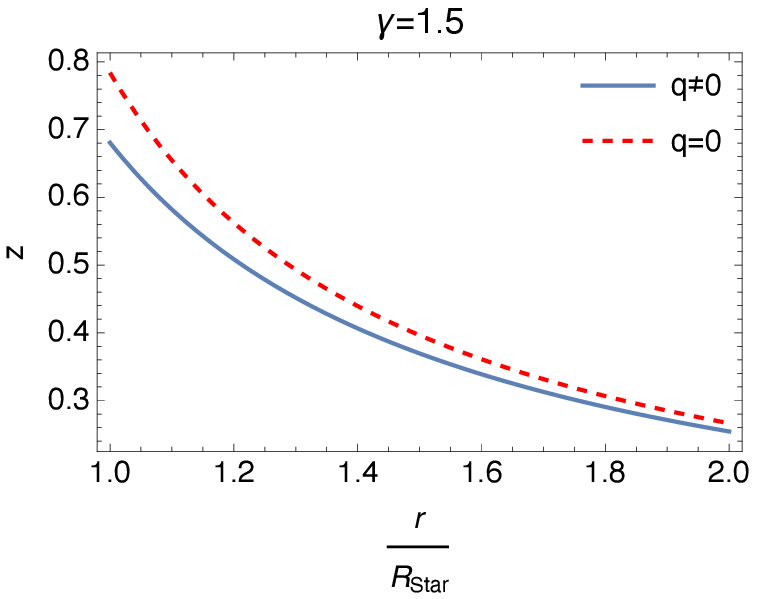}} &

  \end{tabular}
  \caption{The graphs show the variation in the redshift $z$ versus $\frac{r}{R_{Star}}$ for the compact object VelaX-1. The plots are generated for static charged $(q\neq0)$  and static uncharged $(q=0)$  cases. Variation for each case are shown on the same graph for different values of deformation parameter $\gamma$.}
\end{figure}

\begin{figure}[H]
\centering
  \begin{tabular}{@{}cccc@{}}
    \includegraphics{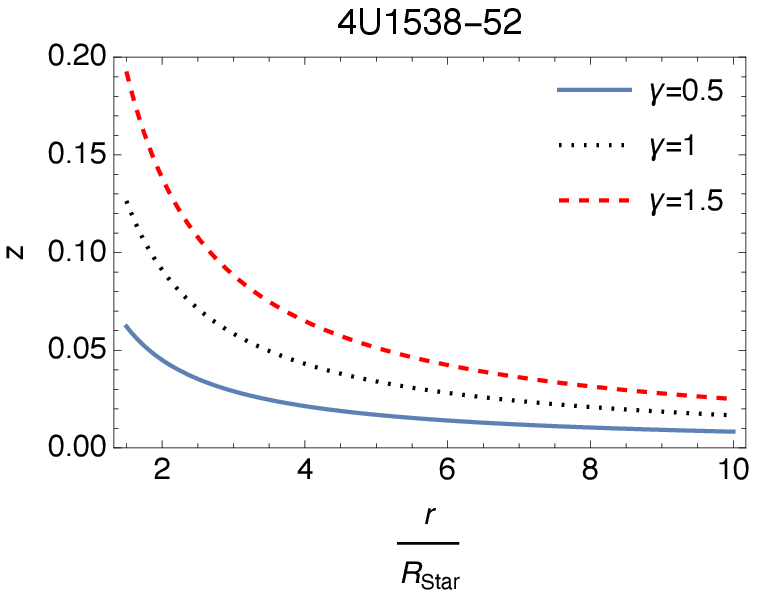} &
    \includegraphics{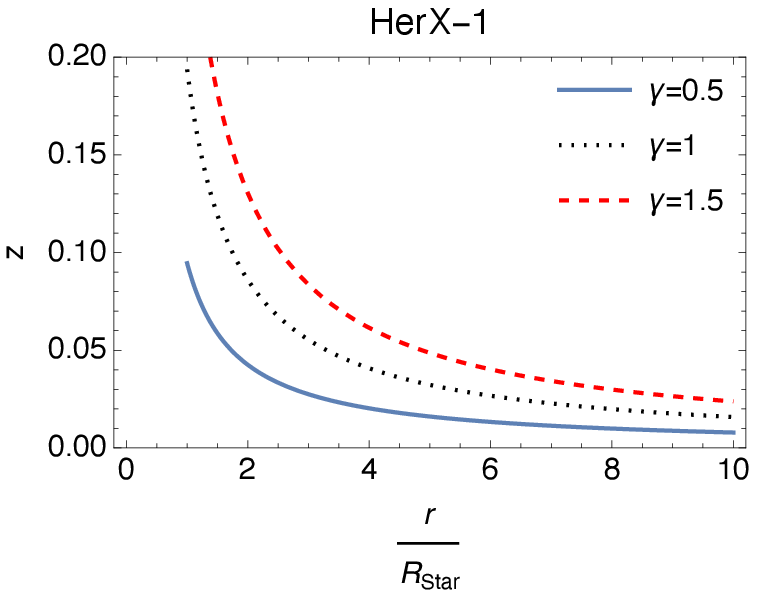} &
                                                         \\
    \includegraphics{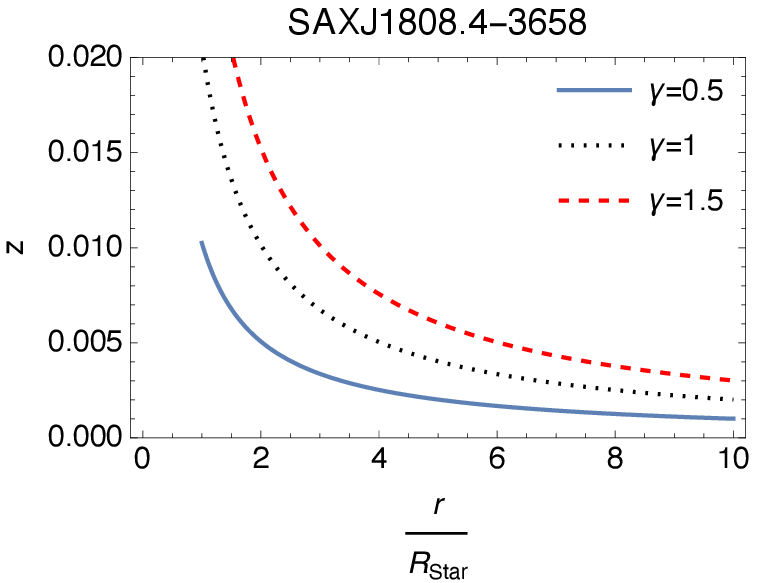} &
    \includegraphics{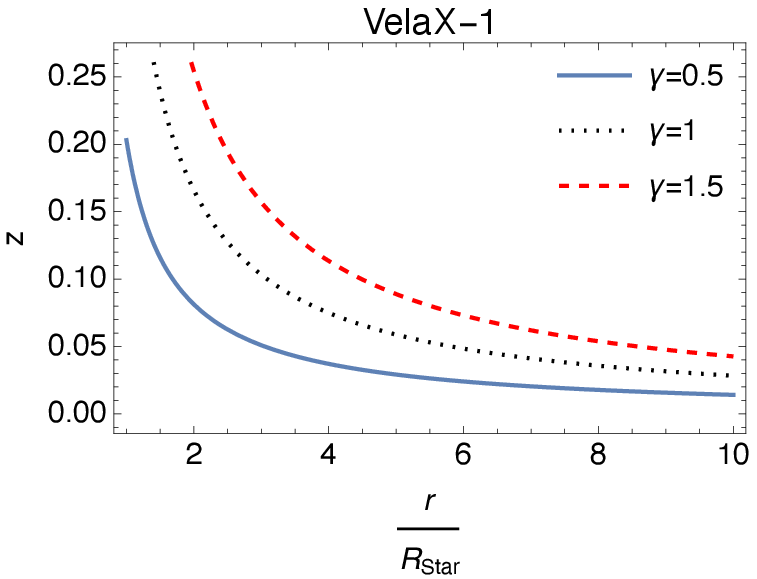} &
                                                            \\
    \end{tabular}
  \caption{Overlapping plots indicating variation in the redshift $z$ against $\frac{r}{R_{Star}}$ for the compact objects given in Table 1 are plotted for comparison. Each plot displays the effect of the deformation parameter $\gamma$ on the redshift $z$. Plots are generated for the charged ZV case for a particular value of $p=q=\frac{1}{\sqrt{2}}$. }
\end{figure}

\section{Conclusion}\label{sect7}

 In this paper, by resorting to the Ernst formalism of the colliding $EM$ spacetime, we charged the $ZV$ metric.We comment that a similar study can also be extended to the stationary $ZV$ metrics \cite{18}. The class of $ZV-$ metrics is known to emerge from interaction of aligned, static rods \cite{3}. Our approach goes beyond the finite rods to the infinite plane waves. In that limit due to symmetry, it becomes possible to find further exact solutions, all by making use of the power of the Ernst formalism. It should  also be remarked that this kind of relation between the $2-$dimensional colliding wave spacetime and the $3-$ dimensional spherical coordinates is nothing but a holographic manifestation. We remark that the infinite class of EM solutions with the second polarization \cite{30} awaits also to be transformed to the ZV space, as done in the present paper, which will involve spinning of the source as well.

 As long as the non-spherical, charged compact objects are concerned, the obtained solution (18) will have astrophysical applications. Observable universe reveals that most planets and stellar objects have magnetic fields. Our Earth has a rather weak field, of the order $B\sim0.5 \; G$, which is yet crucial for life on Earth. Most of the other planets also have magnetic fields, stronger or weaker than that of Earth. Gravastars, however, have extremely high fields of the order $\sim 10^{14}\; G$. When the existence of these magnetic fields is combined with the non-spherical topology of planets/stars, we encounter automatically with the case of the charged $ZV$ metric. This has three parameters, $m$, $q$ and $\gamma$. For $\gamma>1$, the object is oblate and for  $\gamma<1$, it  is prolate. Obviously, $\gamma=1$ gives the spherically symmetric Reissner-Nordstrom $(RN)$ solution and its uniqueness dictates that any other class of $EM$ solutions must all agree at $\gamma=1$.  The source of our metric can both be  pure electric and pure magnetic. Test particle have been investigated through the effective potential for both cases. Their combination, which contains both electric and magnetic fields, sounds more realistic and it can be considered as a separate study. In the light of the existence of directional singularities, i.e. dependence on the angle $\theta$, the singularity structure can further be investigated by the quantum probes \cite{16}. The integrability or non-integrability of the geodesics, chaotic behaviours \cite{17} and the analysis of the outermost singularity $r_{\Delta}=m(1+p)$ can further be analysed. As an application of the found charged $ZV$ metric in this article, for limited parameter $\gamma$, we have shown how the lensing becomes by using the $GB$ theorem. In order to be able to apply the $GB$ theorem to the case of $\theta=\pi/2$, we consider the class of linearzed geodesics near to the equilateral plane. With this restriction geodesics that normally leave the plane will be projected to the same plane. The same is carried out also for the stationary ZV spacetime.

\textbf{Acknowledgment:}

We extend our  sincere gratitude to İ. Gullu, S. H. Mazharimousavi and H. Gursel for their helpful comments.

\end{document}